\documentclass[pdflatex,sn-mathphys-num]{sn-jnl}

\usepackage{xr}
\usepackage{graphicx}%
\usepackage{multirow}%
\usepackage{amsmath,amssymb,amsfonts}%
\usepackage{amsthm}%
\usepackage{mathrsfs}%
\usepackage[title]{appendix}%
\usepackage{xcolor}%
\usepackage{textcomp}%
\usepackage{manyfoot}%
\usepackage{booktabs}%
\usepackage{algorithm}%
\usepackage{algorithmicx}%
\usepackage{algpseudocode}%
\usepackage{listings}%
\usepackage{lmodern}

\begin{document}

\title[Article Title]{High-fidelity microsecond-scale cellular imaging using two-axis compressed streak imaging fluorescence microscopy}


\author*[1,2]{\fnm{Mark A.} \sur{Keppler}}\email{keppler.mark@gmail.com}

\author[1]{\fnm{Sean P.} \sur{O'Connor}}
\author[3]{\fnm{Zachary A.} \sur{Steelman}}
\author[4]{\fnm{Xianglei} \sur{Liu}}
\author[4]{\fnm{Jinyang} \sur{Liang}}
\author[2]{\fnm{Vladislav V.} \sur{Yakovlev}}
\author*[2,3]{\fnm{Joel N.} \sur{Bixler}}\email{joel.bixler.1@us.af.mil}

\affil[1]{\orgname{SAIC}, \orgaddress{\street{4141 Petroleum Dr}, \city{JBSA Fort Sam Houston}, \postcode{78234}, \state{Texas}, \country{USA}}}

\affil[2]{\orgdiv{Department of Biomedical Engineering}, \orgname{Texas A\&M University}, \orgaddress{\street{400 Bizzell St}, \city{College Station}, \postcode{77843}, \state{Texas}, \country{USA}}}

\affil[3]{\orgname{Air Force Research Laboratory}, \orgaddress{\street{4141 Petroleum Dr}, \city{JBSA Fort Sam Houston}, \postcode{78234}, \state{Texas}, \country{USA}}}

\affil[4]{\orgdiv{Centre Énergie Matériaux Télécommunications}, \orgname{Institut National de la Recherche Scientifique}, \orgaddress{\street{1650 Bd Lionel-Boulet}, \city{Varennes}, \postcode{J3X1P7}, \state{Québec}, \country{Canada}}}


\abstract{Compressed streak imaging (CSI) is a computational imaging strategy that can acquire video at over 150 trillion frames per second. Despite this achievement, CSI faces challenges in detecting subtle intensity fluctuations in slow-moving, continuously illuminated objects. This limitation, largely attributable to high streak compression and motion blur, has curtailed the broader adoption of CSI in cellular fluorescence microscopy. To address these issues and expand the utility of CSI, we developed a two-axis compressed streak imaging (TACSI) method that results in significant improvements to the reconstructed video fidelity. TACSI introduces a second scanning axis which shuttles a conjugate image of the object with respect to the coded aperture. The moving image decreases the streak compression ratio and produces a “flash and shutter” phenomenon that reduces coded aperture motion blur, overcoming the limitations of current CSI technologies. This approach is supported with an analytical model describing the TACSI compression ratio, along with simulated and empirical measurements. We demonstrate TACSI’s ability to measure rapid variations in cell membrane potentials, previously unattainable with conventional CSI. This method has broad implications for high-speed photography, including visualization of action potentials, muscle contractions, and enzymatic reactions that occur on microsecond and faster timescales using fluorescence microscopy.}




\maketitle

\section{Introduction}\label{sec:introduction}

Fluorescence microscopy is one of the most important tools driving biological discovery. Over the past several decades, the expansion of fluorescent labels and functional proteins has revolutionized our understanding of biological and biochemical interactions across a wide range of systems and scales. However, many important biological processes occur on sub-millisecond time frames, beyond the temporal resolution of conventional video-rate imaging. Such activities include ion channel kinetics\cite{Beier2012,Maffeo2012}, action potentials\cite{Yang2018}, vesicle trafficking\cite{Joselevitch2020}, calcium signaling from synaptic transmission and muscle cell contraction\cite{Berridge2003,Edwards2012}, and molecular motor activity\cite{Kaya2017}. With sufficient temporal resolution, it also becomes possible to visualize fluorescence lifetimes\cite{Shirshin2022} and phosphorescence lifetimes\cite{Wang2020}, enabling the monitoring of an exceptionally wide variety of processes in cell and tissue biology, including metabolic activity, drug efficacy, and disease progression\cite{Datta2020}. 

Even with highly efficient fluorophores and sensitive back-illuminated camera sensors, observing rapid biological events using fluorescence microscopy is still challenging at sub-millisecond timescales due to read noise and dynamic range constraints\cite{Sanderson2014,Stelzer1998}. Lower photon counts at the increased frame rates can result in read-noise limited measurements. Modern scientific cameras can achieve a single-photon read noise performance of $<$0.3 e$^{-}$ RMS\cite{Nguyen2018,Teranishi2012}; however, even the most state-of-the-art ultra-high-speed (UHS) imaging sensors exhibit read noise at least 17-fold higher\cite{Yue2023}. This severely limits the applicability of UHS sensors to problems in cellular biology, where signal is also bounded to avoid phototoxicity and photobleaching. Given the highly non-deterministic nature of most biological processes, there is a critical need for imaging systems that combine high spatial and temporal resolution with low read-noise.

Over the past decade there has been a proliferation of single-axis compressed streak imaging (CSI) systems, including compressed ultrafast photography (CUP)\cite{Gao2014,Liu2019a}, which have the ability to reconstruct a video from a single snap-shot image. This enables UHS imaging using low read noise and high dynamic range scientific complementary metal-oxide-semiconductor (sCMOS) and electron multiplying charge-coupled device (EMCCD) cameras\cite{Qi2020}. The development of CSI can be attributed to compressed sampling theory\cite{Donoho2006,Candes2006}. Compressed sampling theory provides minimum sufficient criteria for recovering high-dimensional signals when sampled below the Nyquist criterion\cite{Jerri1977}. For a signal to be recovered after compressed sampling, it should have a sparse representation in a basis and be incoherently sampled\cite{Baraniuk2007,Candes2008}. The primary aim of compressed sampling is to collect the minimum feature set required to recover a high resolution signal from an under-sampled measurement. Recent advances in compressed reconstruction algorithms\cite{Ma2021a,Bioucas-Dias2007,Boyd2010} have led to the development of a general CSI framework\cite{Liang2024} which can acquire ultra-high-speed video with up to femtosecond temporal resolution from a single streak image using a scientific camera\cite{Liu2024}.

Conventional CUP uses a streak camera to achieve sub-nanosecond temporal resolution via electronic scanning\cite{Gao2014}. Photons must first be converted into electrons by a photocathode before translating them with a time varying electric field. The electrons are then converted back into photons at a phosphor screen that is then imaged by a scientfic camera. Unfortunately, high dynamic range streak cameras equipped with microchannel plates to amplify low-light signals can cost over \$150,000 USD, and cannot capture events longer than 1 to 10 ms. By comparison, CSI using mechanical scanning elements can achieve microsecond temporal resolution, with the duration limited only by the camera integration time, for under \$30,000 USD using low-cost galvo-scanners\cite{Liu2019a,Matin2021,Keppler2022,Wang2020}. At these time-scales, galvo-scanner systems can achieve continuous acquisition of high temporal resolution kymograms at up to 1.5 MFPS\cite{Liu2019a}.

While CSI schemes offer a cost-effective path to low read noise and ultrafast imaging, they have shown limited utility for fluorescence-intensity imaging of stationary objects.  Often in cell microscopy, per-pixel signals are only on the order of hundreds of counts (DN) and significant changes fluctuate by $\ll$30\%\cite{Kim2023,Kriszt2017,Dana2016}. Applying CSI, Ma et al. reconstructed fluorescence-lifetime videos of cells; however, the signal of interest was the large, monotonic intensity decay following pulsed-laser excitation, not small steady-state intensity changes\cite{Ma2021a}. Because cells remain stationary with respect to the coded aperture (CA) throughout the image acquisition period, they are encoded by the same elements of the CA. Under these conditions, compressed streak images exhibit a high compression ratio and excessive motion blur. Motion blur and compression ratio appear to have previously been mitigated by restricting CSI applications to imaging objects that move at high-speeds and orthogonal to the streak trajectory, such as a pulse of light propagating through free space \cite{Gao2014}, or fluorescent droplets flowing through a micro-channel \cite{Matin2021}. The irradiance from a moving object decreases the exposure duration for each CA element by creating a flash and shutter phenomenon, similar to methods used in late 19$^{th}$ century bullet photography\cite{Boys1893}.

In this work, we have developed a two-axis compressed streak imaging (TACSI) methodology which introduces a second scanning axis for an intermediate image, orthogonal to the conventional streak axis. We validate the hypothesis that translating an intermediate image of a stationary or slow-moving object can reduce video reconstruction artifacts from encoded videos with high streak compression ratios and motion blur under continuous illumination conditions. This approach was investigated by mathematical model and simulation, then validated experimentally. We also demonstrate that the mathematical model of the TACSI compression ratio is a more general form of the equation developed for conventional single-axis CSI\cite{Ma2021a}. We present, for the first time, successful CSI reconstructions of membrane voltage responses from cells loaded with voltage-sensitive dye that are subjected to microsecond-scale pulsed electric fields (PEF) - a stimulus known to induce near instantaneous changes in membrane potential. TACSI was able to reconstruct video of spatial and temporal field effects, where conventional single-axis compressed streak images were impossible to reconstruct. All results were acquired without the use of a second camera\cite{Liang2015}, space and intensity constraints\cite{Zhu2016}, or loss-less encoding\cite{Liang2017}. This work has broad implications for high-speed, read noise-limited photography across a wide range of applications and scales.

\section{Results}\label{sec:results}

\subsection{TACSI fundamentals}\label{sec:results:two_axis_csi_principle}

The following definitions will be used throughout to understand the fundamentals of TACSI. We will refer to conventional CSI as a single-axis technique. The term continuously illuminated refers to the duration of a single compressed streak image acquisition. A stationary or slow-moving object should be considered relative to the translation speed of the CA image at the camera sensor, which provides the constraint for the desired reconstructed video frame rate. Compression ratio is defined as the ratio between the full resolution video and the compressed streak image size, as measured by the number of pixels that convey information about the object under investigation\cite{Ma2021a}. A higher compression ratio increases the probability of reconstruction artifacts and loss of resolution, but allows smaller file sizes and shorter reconstruction times\cite{Gersho1992}. The words object under investigation refer to objects that are smaller than the camera field of view with clearly defined boundaries. This could include individual cells or microbeads against a dark background, but can be extended to scenes with more complex features.

\begin{figure}[ht!]
\centering
\includegraphics[width=0.9\textwidth]{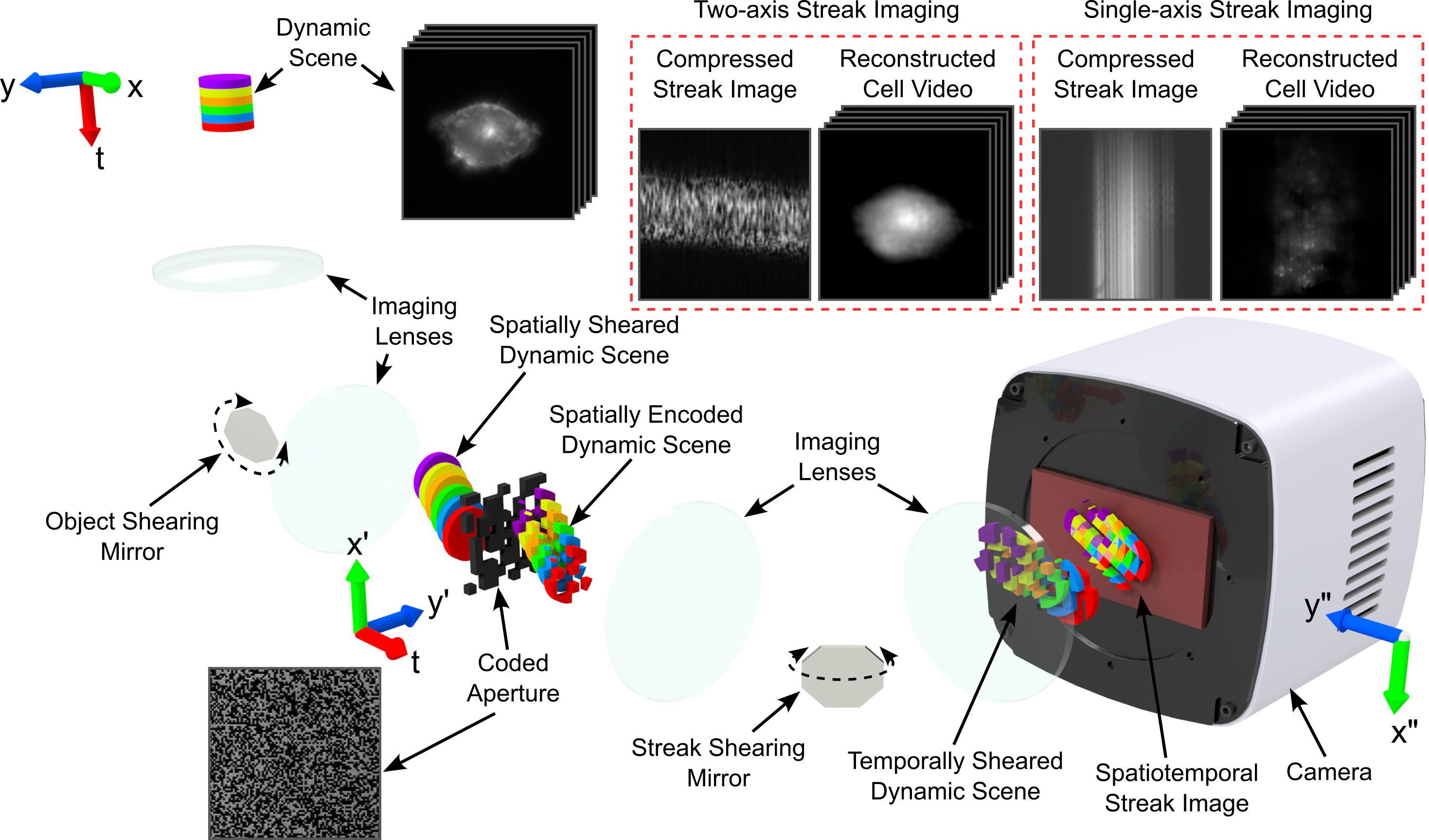}
\caption{Compressed streak imaging conceptual diagram and example images. This diagram shows the encoding scheme required to capture a digital single- or two-axis encoded streak image. Each colored tile represents a time step in the dynamic scene. A representative widefield image of a fluorescent CHO-K1 cell and a brightfield image of the CA are presented as examples. The encoded spatiotemporal streak image of the cell for the single- and two-axis modality can be observed, along with single frames from the reconstructed video. While lenses and mirrors are depicted for clarity, alternative optical elements or arrangements achieving the same conceptual outcome are equally valid.}
\label{fig:conceptual_diagram}
\end{figure}

Figure \ref{fig:conceptual_diagram} shows the components of a TACSI system and their functions. A TACSI system consists of two assemblies: an object shearing relay and a CSI system. A minimal CSI system consists of a CA, streak shearing relay, and a camera. The object and streak shearing relays illustrated in Figure \ref{fig:conceptual_diagram} are both defined by two lenses that relay the images and an oscillating mirror to shear the scene. The results presented here were acquired with conventional inverted fluorescence microscopy; however, the principle could be broadly extended to other forms of photography. As an example, a widefield image of a fluorescent CHO-K1 cell represents the dynamic scene under investigation in the figure below.

The object shearing relay gives the experimenter control of the position and velocity of an object under investigation on a trajectory orthogonal to the CSI system's streak shearing vector. The object shearing relay shears and projects the dynamic scene to a Gaussian distributed pseudo-random CA. The example in Figure \ref{fig:conceptual_diagram} is an opto-mechanical axis, with the object shearing mirror positioned at the Fourier plane of a 4f telescope.

Representative images of the coded aperture (bottom left), along with two-axis and single-axis spatiotemporal streak images (top right) acquired from the CHO-K1 cell depicted in the widefield image are also included in the figure. Here, motion blur completely obscures the pattern of the CA when the object is stationary (single-axis streak image), while individual elements are well defined when the object is moving (two-axis streak image). This demonstrates that the object under investigation can operate as a streaked source, shortening the CA element exposure duration. Please see SI Appendix, Figure \ref{supp:fig:streaked_source_1893}, for a simplified illustration of the physical principle behind streaked source motion blur reduction. It is important to emphasize that time is encoded along the vertical axis of the spatiotemporal streak images. If we consider any fixed vertical distance, the horizontal translation produced by the object shearing relay results in spreading information over a greater surface area of the camera sensor. Translating the object orthogonal to the vertical axis maximizes the surface area, thereby minimizing the compression ratio.

A conventional CSI system forms the streak shearing axis, which temporally shears the spatially encoded dynamic scene, and projects the resultant spatiotemporal streak image to a camera. The CA can be produced with a digital micro-mirror device (DMD), a photo-lithographic or printed mask, or a liquid crystal spatial light modulator (LC-SLM)\cite{Lai2024}. Shearing can be accomplished electronically with a streak tube\cite{Gao2014}, mechanically using a galvo-scanner\cite{Liu2019a} or polygon scanning mirror\cite{Wang2020}, or optically via a chirped pulse\cite{Liu2024}.

The TACSI forward model is a modified version of the CUP forward model\cite{Gao2014}, updated with an additional shearing operation. The shearing operator translates the dynamic scene $I(x,y,t)$ prior to spatial encoding, producing a sheared dynamic scene $I_{s}(x',y',t)$ at the coded aperture plane. This procedure is described by Equation (\ref{eq:intesity_shearing})\cite{Gao2014}:
\begin{equation}
\label{eq:intesity_shearing}
I_{s}(x',y',t)=\textbf{S}_{1}I(x,y,t) \,,
\end{equation}
where $\textbf{S}_{1}$ is the linear object shearing operator. 

The optical energy from the encoded spatiotemporal streak image $E(m,n)$ is then integrated by the camera at each pixel $(m,n)$. This is defined by Equation (\ref{eq:forward_model}):
\begin{equation}
\label{eq:forward_model}
E(m,n)=\textbf{T}\textbf{S}_{2}\textbf{C}I_{s}(x',y',t) \,,
\end{equation}
where $I_{s}(x',y',t)$ is encrypted by a spatial encoding operator $\textbf{C}$, and translated by the temporal shearing operator $\textbf{S}_{2}$, before being projected into pixel-space by the spatiotemporal encoding operator $\textbf{T}$.

An estimate of the sheared dynamic scene $\hat{I_{s}}(x',y',t)$ can be reconstructed from the compressed streak image using a compressed sensing (CS) algorithm such as TwIST\cite{Bioucas-Dias2007} or ADMM-PnP\cite{Boyd2010}. The CS algorithm solves the inverse problem given by Equation (\ref{eq:inverse_model})\cite{Qi2020}:
\begin{equation}
\label{eq:inverse_model}
\hat{I_{s}}(x',y',t)=\min_{I_{s}} \boldsymbol{\Phi}[I_{s}(x',y',t)] \text{ subject to } \textbf{T}\textbf{S}_{2}\textbf{C}I_{s}(x',y',t) \,,
\end{equation}
where $\Phi[I_{s}(x',y',t)]$ represents $I_{s}(x',y',t)$ in a sparse basis.

A dynamic scene estimate $\hat{I}(x,y,t)$ is then reproduced by digitally de-scanning the reconstructed dynamic scene as given by Equation (\ref{eq:intesity_unshearing}):
\begin{equation}
\label{eq:intesity_unshearing}
\hat{I}(x,y,t)=\textbf{S}_{1}^{-1}\hat{I_{s}}(x',y',t) \,.
\end{equation}

As an example, a dynamic scene was acquired from the fluorescence emission from a cell. A widefield image of the cell is shown in Figure \ref{fig:conceptual_diagram}, along with single frames from videos reconstructed using single-axis and two-axis spatiotemporal streak images, respectively. Because it is common to enhance results by overlaying the reconstructed intensity map onto a widefield image\cite{Liang2015}, or using a widefield image to provide space- and intensity constraints for the CS algorithm\cite{Zhu2016}, it is necessary to emphasize that the reconstructed frames displayed in Figure \ref{fig:conceptual_diagram} only show results from the raw reconstructed video. Single-axis streak is unable to reconstruct cellular features, while the TACSI reconstruction captures the shape and intensity profile.

\subsection{TACSI compression ratio}\label{sec:results:compression ratio}

Controlling the portion of the camera sensor over which the digital spatiotemporal streak signal is acquired has an effect on the signal's compression ratio, defined generally as in Equation (\ref{eq:general_compression_ratio})\cite{Ma2021a}:
\begin{equation}
\label{eq:general_compression_ratio}
CR = \frac{N_{x}N_{y}N_{t}}{N_{s'}},
\end{equation}
where $N_{x}$, $N_{y}$, and $N_{t}$ are the number of pixels required to convey the spatial information along the x and y coordinate axes and the time information of a full resolution video. $N_{s'}$ refers to the number of spatial pixels after the video has been compressed into a single streak image. It is important to note that compression ratio treats the extent of the signal and not that of the camera sensor. This is further clarified in Section \ref{sec:results:compression ratio_comparison} with the definition of sequence depth limits.

The description of compression ratio given by Equation (\ref{eq:general_compression_ratio}) can be extended to describe TACSI by considering the model in SI Appendix, Figure \ref{supp:fig:compression_ratio_model}, which depicts the path traced by an object as defined by the object's velocity and the streak velocity. Streak velocity $v_{s}$ is used to refer to the translation rate of the coded aperture image formed at the camera sensor along the sensor's vertical axis. The term object velocity $v_{o}$ is used to describe the speed and direction of motion of the image of the object under investigation at the camera sensor plane, with coordinates defined in relation to the horizontal and vertical sensor axes. An amorphously shaped object is chosen to generalize the model because many scenes will not present with a rectangular geometry, such as an isolated cell. Simplifying the model using a bounding box around the object will yield an overestimate of the compression ratio in this case.

The post-reconstruction frame rate provides a constraint for determining the number of frames in the high resolution video. This is necessary because under empirical conditions it is generally not possible to acquire a full resolution image for comparison. The frame rate of a compressed streak imaging system is defined in Equation (\ref{eq:streak_fps})\cite{Gao2014}:
\begin{equation}
\label{eq:streak_fps}
FPS=\frac{v_{s}}{p} \,,
\end{equation}
where $v_{s}$ is the streak velocity and $p$ is the pixel pitch of the camera. 

The surface area of the streak in square pixels can be determined from Equation (\ref{eq:streak_surface}):
\begin{equation}
\label{eq:streak_surface}
N_{s'} = \frac{w_{\perp}t\sqrt{v_{o}^{2}+v_{s}^{2}}+s_{o}}{p^{2}}\,,
\end{equation}
where $s_{o}$ is the surface area of the object in square meters, $w_{\perp}$ is the width of the object along an axis perpendicular to its trajectory, and $t$ is the streak duration over which the object overlaps with itself. The number of image frames in the high resolution video can be found using Equation (\ref{eq:num_frames}):
\begin{equation}
\label{eq:num_frames}
N_{t}=\frac{v_{s}t}{p} \,,
\end{equation}
and the number of spatial pixels as in Equation (\ref{eq:full_surface}):
\begin{equation}
\label{eq:full_surface}
N_{s}=\frac{s_{o}}{p^{2}} \,.
\end{equation}

Equation (\ref{eq:specific_compression_ratio}) can be used to estimate the compression ratio of the streak image:
\begin{equation}
\label{eq:specific_compression_ratio}
CR=\frac{N_{s}N_{t}}{N_{s'}}=\frac{s_{o}|v_{s}|t}{p(w_{\perp}t\sqrt{v_{o}^{2}+v_{s}^{2}}+s_{o})} \,.
\end{equation}

Note that the compression ratio only increases over the time interval where the object under investigation overlaps with itself, while streaked across the camera sensor. For a circular object, this would be the time required for the object to travel twice its diameter. Under physically realizable conditions, all variables except the object and streak speeds remain positive and real valued. A decreasing compression ratio is predicted by an increasing object velocity, which supports the hypothesis. The trend in compression ratio as the object rate to streak rate ratio increases is shown in SI Appendix, Figure \ref{supp:fig:compression_ratio_plots}.

\subsection{Comparing TACSI and single-axis compression ratios}\label{sec:results:compression ratio_comparison}

The compression ratio of a single-axis compressed streak image can be calculated using Equation (\ref{eq:compression_ratio_1D})\cite{Ma2021a}:
\begin{equation}
\label{eq:compression_ratio_1D}
CR_{1D} = \frac{N_{x}N_{y}N_{t}}{N_{x}(N_{y}+N_{t})}.
\end{equation}
The single-axis compression ratio in Equation (\ref{eq:compression_ratio_1D}) has been slightly modified from the form produced by Ma et al. This version conveys the maximum length of a streak that can be acquired by a sensor with $P_{y}$ pixels, whereas the version derived by Ma et al. calculated the length of the streak produced by a video with $N_{t}$ frames. This distinction allows the TACSI compression ratio, given by Equation (\ref{eq:specific_compression_ratio}), to be directly compared to Equation (\ref{eq:compression_ratio_1D}) after discretizing the variables and adopting a rectangular geometry. For convenience, the ratio between the object and streak speed is defined as the streak ratio $r$:
\begin{equation}
\label{eq:streak_ratio}
r = \frac{v_{o}}{v_{s}}.
\end{equation}
The discrete form of the TACSI compression ratio can be described by:
\begin{equation}
\label{eq:compression_ratio_2D}
CR_{2D} = \frac{N_{x}N_{y}N_{t}}{N_{t}\biggl[N_{x}sin\biggl(tan^{-1}\biggl(\frac{1}{r}\biggl)\biggl) + N_{y}cos\biggl(tan^{-1}\biggl(\frac{1}{r}\biggl)\biggl)\biggl]\sqrt{r^{2}+1}+N_{x}N_{y}}.
\end{equation}
A complete derivation of Equation (\ref{eq:compression_ratio_2D}) is included in the supplementary material, along with a conceptual diagram in SI Appendix, Figure \ref{supp:fig:discrete_cr}. In the limit as the streak ratio approaches zero, the TACSI compression ratio reduces to the single-axis case.

The maximum sequence depth for a single-axis streak video is limited by the height of the scene and the number of row pixels:
\begin{equation}
\label{eq:single-axis_frame_limit}
N_{t} \leq P_{y}-N_{y},
\end{equation}
TACSI requires considering constraints along the horizontal and vertical axis of the imaging sensor:
\begin{equation}
\label{eq:two-axis_frame_limit}
N_{t} \leq min \biggl\{ P_{y}-N_{y},\frac{P_{x}-N_{x}}{r} \biggl\}.
\end{equation}
As mentioned in Section \ref{sec:results:compression ratio}, these definitions of sequence depth assume that the spatial extent of the scene is smaller than the extent of the camera sensor. While it is possible to extend a scene beyond the bounds of the camera sensor, this would result in cropping the field of view and a higher than necessary compression ratio. As was the case in Section \ref{sec:results:compression ratio}, the compression ratio only increases for the local sequence of frames that convolve together.

\begin{figure}[ht!]
\centering
\includegraphics[width=0.9\textwidth]{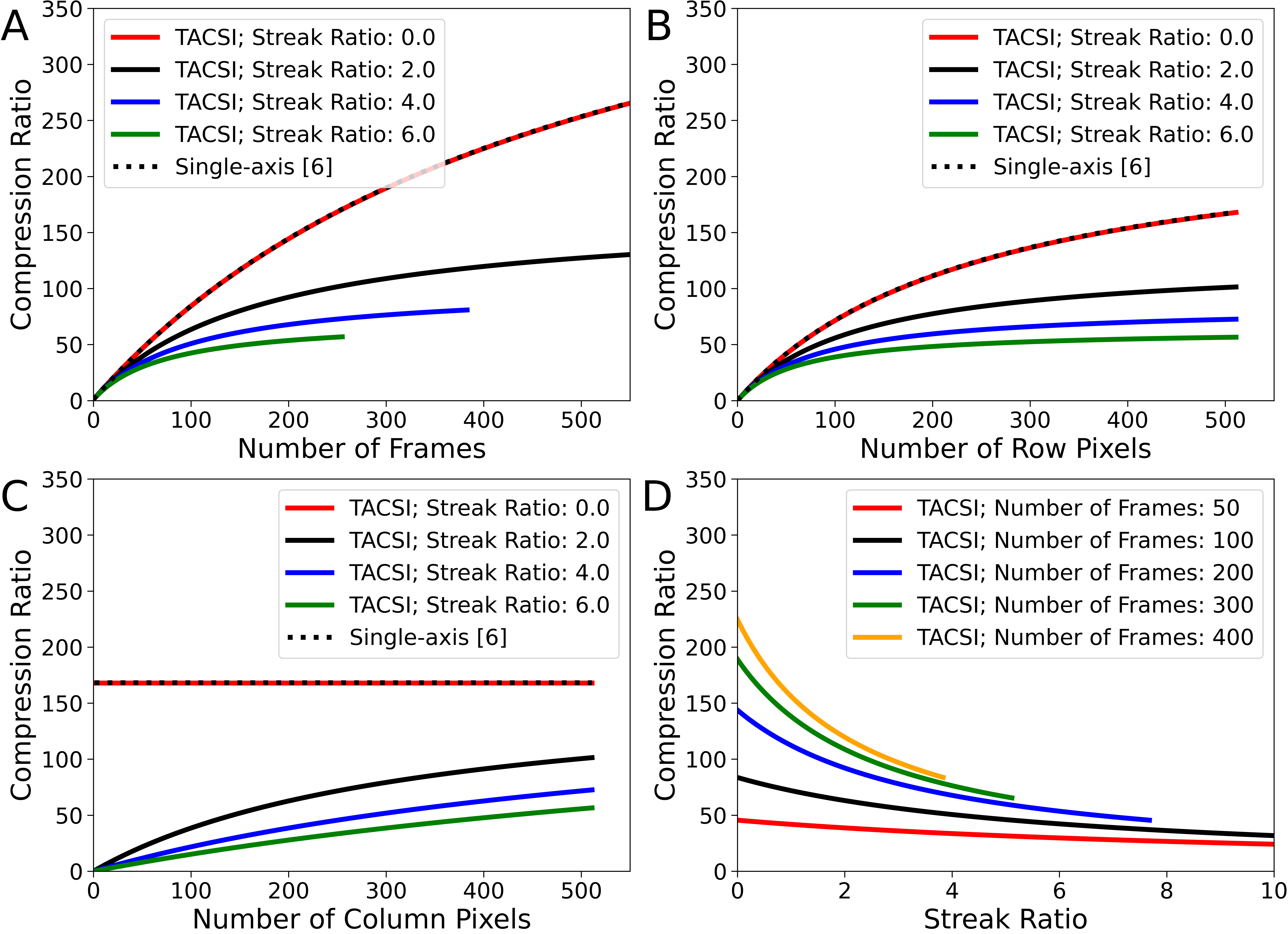}
\caption{Comparison of compression ratios between single-axis and TACSI streak imaging systems. (A) Compression ratio calculated for a fixed 512 by 512 pixel scene acquired on a 2048 by 2048 pixel sensor, with varying numbers of frames. The streak trajectory is oriented vertically (y-axis), and the object trajectory horizontally (x-axis). (B) Compression ratio calculated with varying numbers of row pixels spanning the rectangular scene, while keeping the number of column pixels fixed at $N_{x}=512$ and the number of frames at $N_{t}=250$. (C) Compression ratio calculated with varying numbers of column pixels spanning the scene, while keeping the number of row pixels constant at $N_{y}=512$ and the number of frames at $N_{t}=250$. (D) Compression ratio as a function of streak ratio for different numbers of frames, calculated for a fixed 512 by 512 pixel scene. The maximum recoverable frames are limited by Equation (\ref{eq:two-axis_frame_limit}).}
\label{fig:streak_ratio_comparison}
\end{figure}

Figure \ref{fig:streak_ratio_comparison} compares the single-axis and TACSI compression ratios. The number of frames and the streak ratio were constrained to a 2048 × 2048 pixel camera sensor using Equation (\ref{eq:two-axis_frame_limit}), which can be observed in Figures \ref{fig:streak_ratio_comparison}A and \ref{fig:streak_ratio_comparison}D. The streak trajectory is oriented vertically along the y-axis, whereas the object's trajectory is horizontal along the x-axis. Equations (\ref{eq:compression_ratio_1D}) and (\ref{eq:compression_ratio_2D}) converge when the scene is stationary ($r=0$). Figure \ref{fig:streak_ratio_comparison}A presents the compression ratio calculated for a 512 × 512 pixel scene. As the streak ratio increases, the compression ratio correspondingly decreases. In Figure \ref{fig:streak_ratio_comparison}B, the compression ratio is calculated by varying the number of row pixels ($N_{y}$) spanning the rectangular scene, while maintaining a constant number of column pixels ($N_{x}=512$) and frames ($N_{t}=250$). Figure \ref{fig:streak_ratio_comparison}C shows the compression ratio when varying the number of column pixels ($N_{x}$) across the scene, keeping the number of row pixels fixed ($N_{y}=512$) and the number of frames constant ($N_{t}=250$). As expected, when the streak ratio is zero, the compression ratio remains unaffected by changes in the number of column pixels. Figure \ref{fig:streak_ratio_comparison}D illustrates the effect on the compression ratio as the number of frames increases for a fixed 512 × 512 pixel scene.

\subsection{Comparing TACSI to conventional CSI}\label{sec:results:two_axis_simulation}

High resolution simulated videos of 10 $\mu m$ fluorescent microbeads with 10 kHz intensity modulations were developed to compare the single- and two-axis CSI approaches without noise. Compressed streak images were simulated from the high resolution videos of stationary beads and beads moving orthogonal to the streak shearing vector at 10-times the shearing rate. The video reconstructions were then compared to the high definition videos to assess fidelity. 

Both single- and two-axis simulations were generated from the same video to keep energy constant. The bead intensity in the videos was chosen such that the maximum intensity in the single-axis streak image was at the 16-bit saturation point of 65,535. The maximum intensity observed in the two-axis streak image was much lower due to the reduced streak compression. 

\begin{figure}[ht!]
\centering
\includegraphics[width=0.9\textwidth]{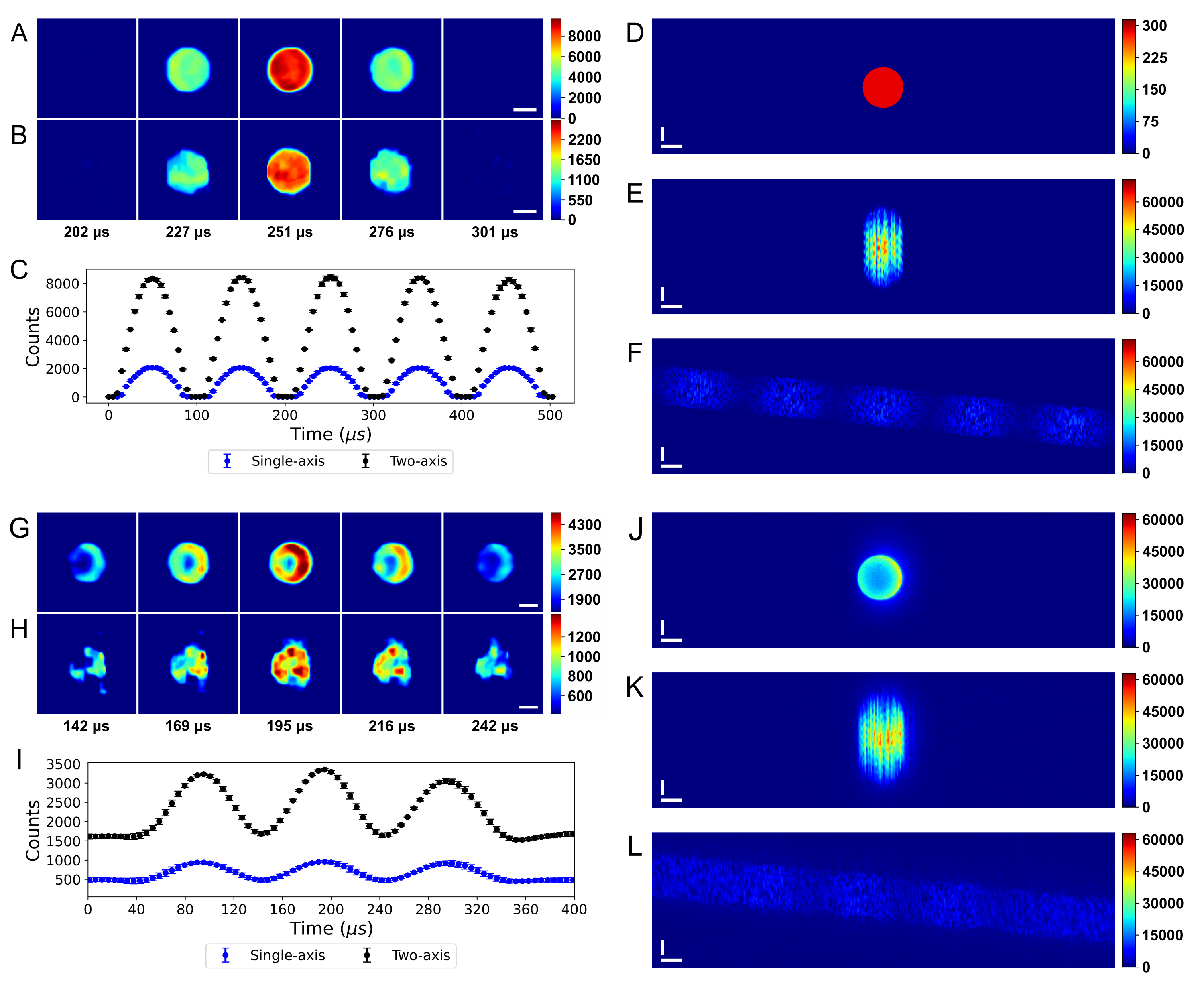}
\caption{Comparison between single- and two-axis compressed video reconstructions of simulated and empirically measured 10 $\mu m$ fluorescent microbeads with a 10 kHz intensity modulation. Sub-figures (A) and (B) compare several frames from the simulated two-axis and single-axis video reconstructions spanning 1 period of modulation, respectively. Sub-figure (C) shows the temporal profile of the average simulated microbead intensity within a 10 $\mu m$ circular region centered on the bead. The error bars are in standard deviations. A single frame from the high resolution video simulation can be observed in sub-figure (D), and sub-figures (E) and (F) show the single- and two-axis simulated streak images, respectively. Sub-figures (G) and (H) show several reconstructed video frames from the empirically measured fluorescent microbeads spanning 1 period of modulation for two-axis and single-axis streak, respectively. Sub-figure (I) shows the average measured intensity within a circular ROI with a 10 $\mu m$ diameter centered on the beads (N=9). Error bars are in standard deviations. Sub-figures (J-L) show a widefield, single-axis streak, and two-axis measured streak image, respectively, containing the same microbead seen in sub-figures (G) and (H). White horizontal scale bars represent 5 $\mu$m, and vertical scale bars represent 100 $\mu s$.}
\label{fig:fundamental_concept}
\end{figure}

Figures \ref{fig:fundamental_concept}A and \ref{fig:fundamental_concept}B show several frames from the two- and single-axis video reconstructions, respectively. These images detail one modulation cycle of the simulated laser, where the illumination intensity ranges from 0 to 100\% relative power. The image frames at 202 and 301 $\mu s$ occur where the modulation intensity is zero. The temporal profiles of the videos reconstructed with the ADMM-PnP algorithm are shown in Figure \ref{fig:fundamental_concept}C. There is a 4.1-fold increase in the bit utilization ratio (BUR) for two-axis counts output by ADMM, while the number of resolvable intensity steps are nearly identical for reconstructions after min/max normalization. The BUR is defined as the ratio between the bit levels spanning the digitized signal to the number of available bit levels. Figure \ref{fig:fundamental_concept}D shows a single frame from the simulated high resolution fluorescent microbead video, while Figures \ref{fig:fundamental_concept}E and \ref{fig:fundamental_concept}F show the single- and two-axis streak images, respectively. The maximum counts in the two-axis streak image is 14,802. This is 4.4 times lower than the single-axis maximum, and implies that two-axis excitation intensity could be increased more than 70\%. It is important to note here that temporal information is encoded along the vertical axis in both single- and two-axis CSI. As such, the apparent transverse streak spreads spatial information without changing the reconstructed frame rate.

\subsection{Intensity modulated fluorescent microbeads}\label{sec:results:fluorescent_beads}

Figures \ref{fig:fundamental_concept}G-\ref{fig:fundamental_concept}L demonstrate that the enhanced BUR of the two-axis technique over conventional single-axis CSI also applies to empirical measurements. Figures \ref{fig:fundamental_concept}G and \ref{fig:fundamental_concept}H show a 10 $\mu$m fluorescent bead imaged for 500 $\mu$s acquired using a two- and single-axis streak with matched effective frame rates, respectively. The modulation intensity was adjusted to produce a fluorescence change in the bead of 100\%. The image frames displayed correspond to the maxima, mid-points, and minima over a single period of the 10 kHz modulation signal. Identical laser power was used for both acquisitions. Figure \ref{fig:fundamental_concept}I shows the temporal profiles as average counts, calculated from 10 $\mu m$ circular regions centered on the beads. While both profiles demonstrate a 10 kHz modulation, the experimental two-axis reconstruction results in a 3.3-fold increase in the BUR ratio over single-axis, with approximately 500 counts on average for single-axis vs 1630 counts for two-axis on a 16-bit scale. Supplemental Video 1 and 2 show the reconstructed single- and two-axis empirically measured fluorescent microbead videos, respectively. SI Appendix, Figure \ref{supp:fig:bead_simulation_comparison}, compares frames from simulated single- and two-axis bead reconstructions to their ground truth. The peak signal to noise ratio (PSNR), structured similarity index measure (SSIM), BUR, and compression ratio (CR) for each method can be found in SI Appendix, Table \ref{supp:tab:microbead_metrics}. The PSNR and SSIM increased by 15\% and 41\%, respectively, as the compression ratio was reduced by roughly 10-fold.

In addition to the increased BUR, TACSI allows for the use of a far greater peak laser power before pixel saturation. With a fluence of around 0.1-0.4 $kW/cm^{2}$, the maximum counts within the single-axis streak reached 98\% of the available well depth. The large irradiance range was influenced by the coded aperture and the fluorescence intensity of the bead. TACSI by comparison only utilized 30\% of the available well depth. This technique provided an improved reconstruction while utilizing a fraction of the sensor's well depth, leaving room for further improvement through increased illumination intensity, depending upon the application.

\begin{figure}[ht!]
\centering
\includegraphics[width=0.9\textwidth]{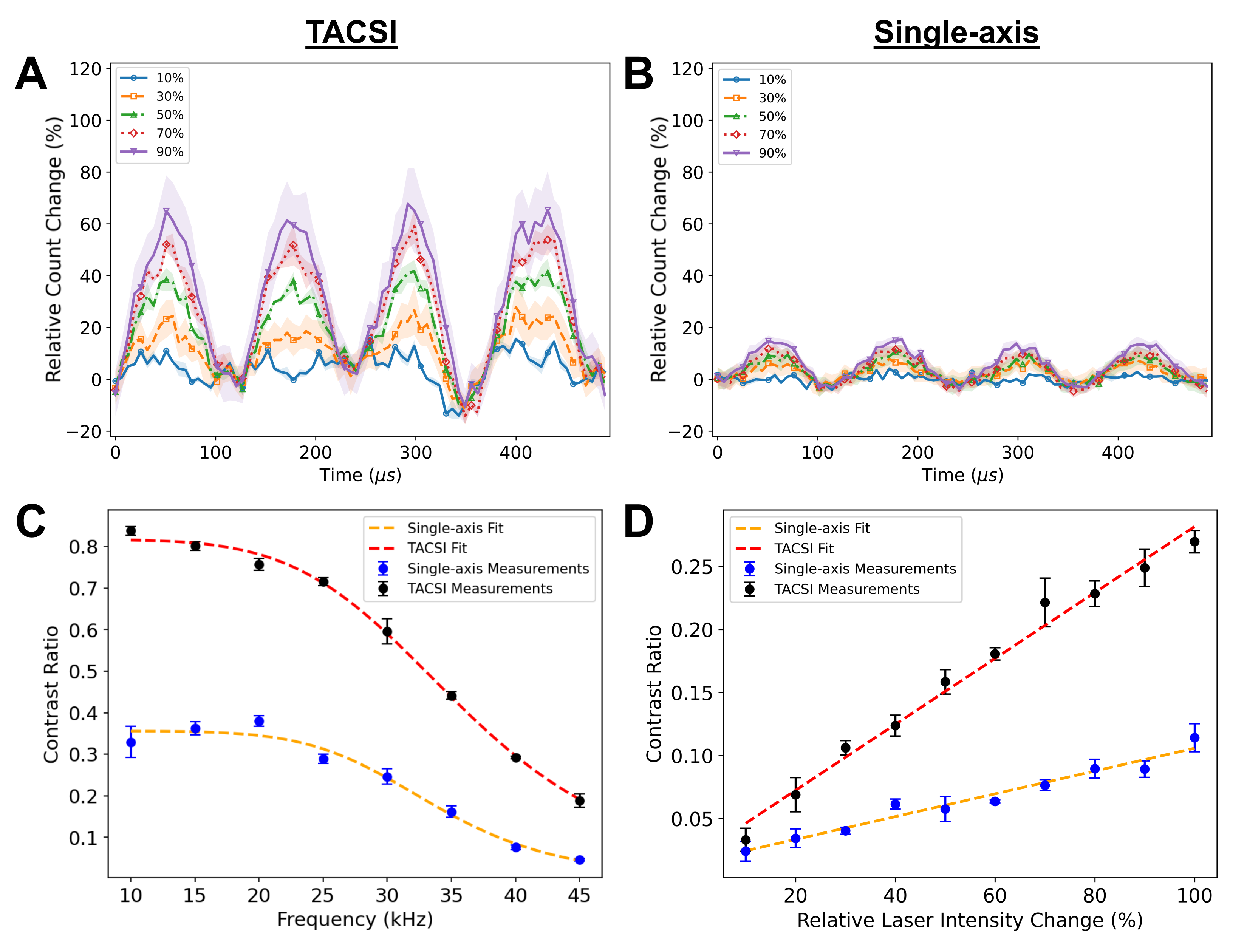}
\caption{Accuracy, temporal resolution, and sensitivity of TACSI versus single-axis CSI. (A) Relative intensity change over time for TACSI from reconstructions of fluorescent micro-beads with sinusoidal modulation amplitude varied from 10-90\%. (B) Same as (A) but for single-axis CSI. (C) Temporal resolution evaluated as fluorescent micro-bead contrast versus laser power modulation frequency. TACSI: black circles ($N=3$) with red dashed fit ($R^2=0.997$, $NRMSE=0.021$); single-axis CSI: blue circles ($N=3$) with orange dashed fit ($R^2=0.98$, $NRMSE=0.077$). TACSI reaches 50\% contrast at 33 $kHz$ (temporal resolution $\sim$30 $\mu s$); single-axis CSI remains below 50\% over the measured range. (D) Sensitivity evaluated as fluorescence micro-bead contrast versus the relative change in the laser power modulation at 10 kHz. TACSI: black circles ($N=3$) with red dashed linear fit (slope = $2.5\times10^{-3}$, $R^2=0.96$, $NRMSE=0.082$); single-axis CSI: blue circles ($N=3$) with orange dashed fit (slope = $9.0\times10^{-4}$, $R^2=0.98$, $NRMSE=0.063$). Error bars represent the standard error of the mean.}
\label{fig:temporal_res_and_sensitivity}
\end{figure}

The relative change in the counts measured from fluorescent microbeads was calculated from reconstructed compressed streak videos. Figure \ref{fig:temporal_res_and_sensitivity}A and \ref{fig:temporal_res_and_sensitivity}B show 4 periods with the bead sinusoidally modulated at 10 kHz ($N=3$). Each colored plot represents the relative change in the min and max laser intensity during modulation, with values from 10\% to 90\%. The mask position was not shifted between replicates, which likely accounts for the large sampling artifact seen at $\sim$360 $\mu s$. This artifact is visible in both the single-axis and TACSI reconstructions. The peak relative change in counts differs from the expected value by roughly 10\% for TACSI, while single-axis differs by up to 75\%. This result correlates well with the reduced compression ratio for TACSI acquisitions. The shaded regions in Figure \ref{fig:temporal_res_and_sensitivity}A and \ref{fig:temporal_res_and_sensitivity}B represent the standard error of the mean.

Figure \ref{fig:temporal_res_and_sensitivity}C displays the contrast for 10 $\mu m$ fluorescent beads as the laser power modulation frequency was swept from 10 to 45 $kHz$. Each data point reflects the mean of 100 pixel-level measurements per bead, averaged over three beads; error bars are the standard error of the mean (SEM). TACSI maintains $\geq$ 50\% contrast down to a $\sim$33 $kHz$ modulation frequency, corresponding to a conservative estimate for the effective temporal resolution at $\sim$30 $\mu s$. Single-axis CSI does not meet the 50\% temporal resolution threshold at any test frequency.

Figure \ref{fig:temporal_res_and_sensitivity}D shows the contrast at a fixed laser power modulation frequency of 10 kHz, while the relative change in the laser intensity was varied from 10\% to 100\%. The 2.7-fold steeper sensitivity slope for TACSI measurements indicates that, for a given percent change in fluorescence, TACSI detects smaller relative intensity changes. 

These data confirm that TACSI simultaneously extends temporal resolution while improving accuracy and sensitivity. The contrast advantage appears to be consistent with the 3.3-fold higher bit-utilization ratio (BUR) observed for TACSI in Figure \ref{fig:fundamental_concept}C and \ref{fig:fundamental_concept}I, underscoring the benefit of distributing the signal over two axes to reduce signal compression.

\subsection{Visualizing high speed cell membrane potential changes}\label{sec:results:cells}

Biological applications in fluorescence microscopy for CSI have remained challenging due to read noise limited signals and the high dynamic range required to capture subtle variations in fluorescence intensity. The increased BUR ratio obtained due to the two-axis technique allows for high fidelity reconstructions of dynamic cellular fluorescence changes with conventional CSI methods. 
\begin{figure}[ht!]
\centering
\includegraphics[width=0.9\textwidth]{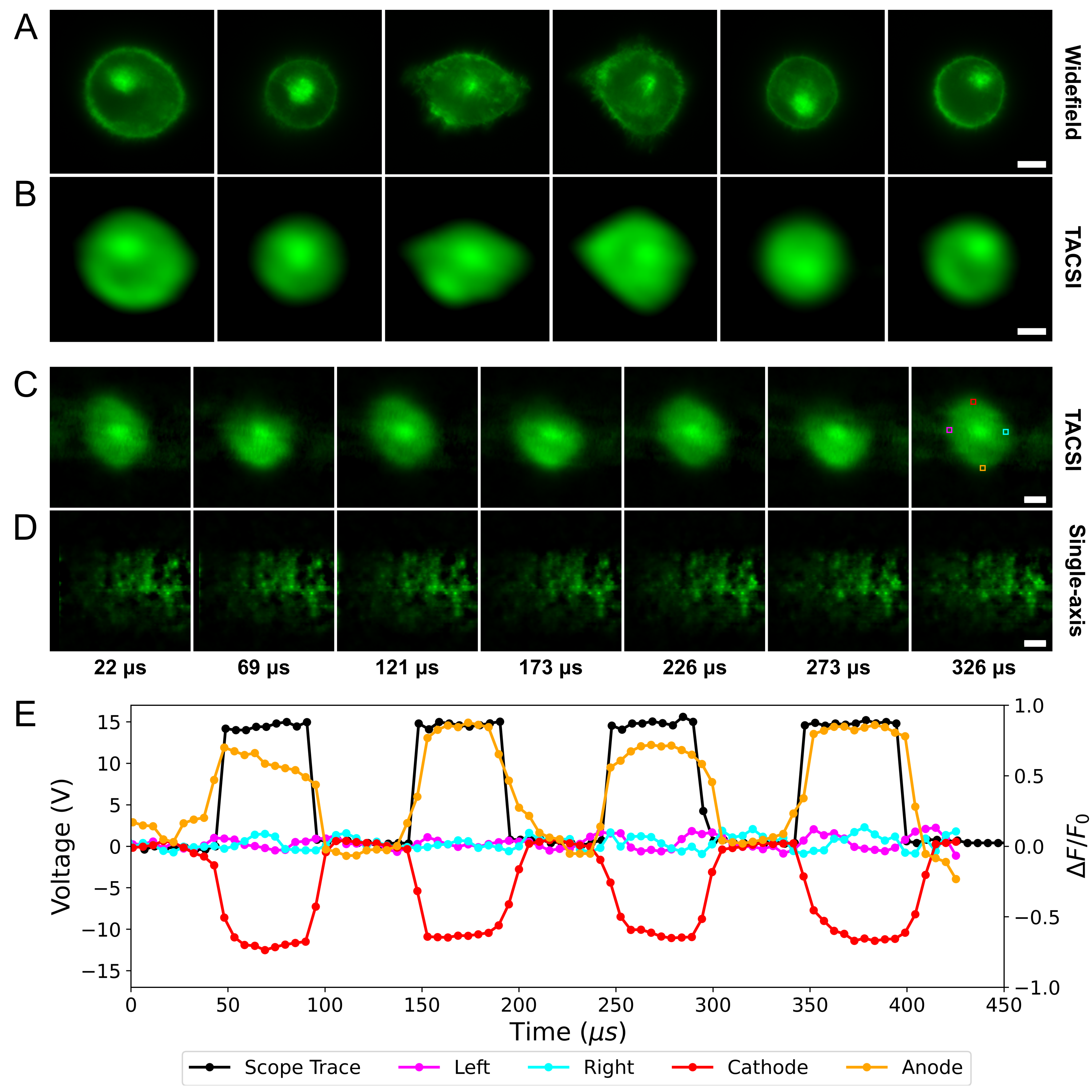}
\caption{High speed video reconstructions showing cell membrane potential responses to PEFs. Comparison between (A) conventional wide-field fluorescence images and (B) averaged two-axis reconstructions of CHO-K1 cells labeled with voltage sensitive dye. Sub-figure (C) presents multiple two-axis reconstructed video frames of CHO-K1 cells during electric pulse delivery, while sub-figure (D) displays several single-axis reconstructed frames of the same process. The temporal profiles of the 4 colored ROIs in last panel of sub-figure (C) can be observed in sub-figure (E) ($n=1$) along with an oscilloscope trace showing the measured voltage at each time point. The red (cathode) and orange (anode) ROIs are positioned at regions of the cell membrane orthogonal to the electric field vector, while the magenta (left) and cyan (right) ROIs are positioned near regions of the membrane parallel to the field vector. The fluorescence response of the membrane at the cathode and anode decreases and increases, respectively. The left and right side of the cell shows no response to the field. The white rectangular scale bars represent 5 $\mu m$.}
\label{fig:cho_cells}
\end{figure}

To validate this concept and further demonstrate the utility of our technique for high-speed fluourescence microscopy applications, images were captured of CHO-K1 cells loaded with FluoVolt\textsuperscript{TM} voltage-sensitive dye while being exposed to PEFs. The response of a CHO-K1 cell to PEFs has been well characterized in literature\cite{Weaver1996,Kotnik1997,Kotnik2000} and this stimulus can be used as a precisely timed event to induce changes in fluorescence intensity\cite{Beier2019}. FluoVolt\textsuperscript{TM} works by photoinduced electron transfer via molecular wire that assembles in the cell membrane such that a green fluorescent protein (GFP) is external to the cell while the quencher is embedded in the membrane\cite{Miller2012}. Electric fields oriented orthogonal to the membrane from outside-to-inside the cell result in electron donation, reducing the GFP fluorescence. Electric fields oriented from inside-to-outside the cell result in an increase in the GFP fluorescence.

Figure \ref{fig:cho_cells} compares video reconstructions of CHO-K1 cells showing the response of FluoVolt\textsuperscript{TM} voltage-sensitive dye to $\sim$ 800 $V/cm$ PEFs with a 50 $\mu$s pulse width and 50\% duty cycle. Streak images were acquired with 500 $\mu$s durations. Frames from the single-axis video reconstructions are seen in Figure \ref{fig:cho_cells}D and primarily consist of reconstruction artifacts with no discernible cellular features. The two-axis reconstructions in Figure \ref{fig:cho_cells}C contain clear cell margins with minimal artifacts. Supplemental Video 3 and 4 show the single- and two-axis reconstructions of the CHO-K1 cells shown above responding to electric pulses, respectively.

Figure \ref{fig:cho_cells}E displays the temporal profile plot from 82 reconstructed frames ($\sim$5.3 $\mu s$ frame interval) of the relative change in FluoVolt\textsuperscript{TM} fluorescence during PEF delivery for the colored square ROIs in Figure \ref{fig:cho_cells}C ($n=1$). Upon applying an electric field, cell membrane regions orthogonal to the field vector exhibit differential fluorescence responses: decreased intensity near the cathode (red ROI) and increased intensity near the anode (orange ROI). The membrane regions parallel to the field vector within the magenta and cyan ROIs show no response. These results are expected based on prior work visualizing membrane charging kinetics due to PEF exposure\cite{Kiester2021}, and demonstrates that TACSI preserves the spatial and temporal features generated by the electric field.

In Figure \ref{fig:cho_cells}A and \ref{fig:cho_cells}B, widefield images of CHO-K1 cells loaded with Fluovolt voltage sensitive dye are compared to averaged two-axis video reconstructions, respectively. The general shape and intensity profile of the cells appears to be preserved, albeit with reduced spatial resolution. The spatial resolution is constrained by the size of the CA elements, with the pixel size representing the ultimate limit.

\section{Discussion}\label{sec:discussion}

\subsection{Enabling Continuous Illumination in CSI} \label{sec:discussion:implications}

TACSI presents a modular solution to simultaneously decrease motion blur and the streak compression ratio, resulting in marked improvements to the temporal resolution, sensitivity, and spatial characteristics of compressed video reconstructions; when the object being investigated is stationary with respect to the CA. Here we have demonstrated that the TACSI compression ratio equation is a generalization that converges to the single-axis equation when the object is stationary. To our knowledge, this is the first demonstration of CSI that can resolve continuously illuminated fluorescently labeled cells and detect subtle variations in their fluorescence intensity. Cell images were acquired with a sequence depth of $\sim$87 frames, which could be increased to 174 frames with 1x1 binning. The sequence depth was counted after removing any frames where the object was partially eclipsed by the edge of the sensor, which varies slightly based on cell size and morphology. If we assume that single-axis CSI reconstructs with comparable video quality at equivalent compression ratios, this would limit the sequence depth to 13 frames (26 frames with 1x1 binning). All attempts to recover video of CHO-K1 cells from single-axis CSI failed to reconstruct any cellular features.

A two-axis galvo-scanner placed at the Fourier plane between the CA and camera would provide a less complex optical system. However, because the object under investigation would not move with respect to the CA, motion blur would be equivalent to the single-axis technique. Frame rate is related to the streak speed, as described by Equation (\ref{eq:streak_fps}). Assuming that it is possible to define frame rate along a diagonal, which is not accounted for by the current model, this approach does not have a mechanism for reducing the streak compression ratio and may introduce interpolation artifacts. The coded aperture moving along a diagonal vector would also result in additional complexity during video reconstruction. Similarly, no advantage can be gained by simply rotating the camera or galvo-scanner, and rotating the camera with respect to the CA would cause a mismatch between the CA elements and the camera pixels. Further details, including simulated compressed streak images for single-axis, the two-axis galvo-scanner positioned between the CA and camera, and TACSI can be found in SI Appendix, Figure \ref{supp:fig:streak_method_comparison}.

Because TACSI does not require modification to existing CS algorithms, it should be possible to implement on multi-dimensional CSI systems for phase sensitivity\cite{Kim2020} or tomography\cite{Lai2021}. Further improvement in image fidelity may be possible via space- and intensity constraints\cite{Zhu2016} and loss-less encoding (LLE) schemes\cite{Liang2017}. While these techniques have been shown to dramatically improve the fidelity of spatial features within compressed streak images, there is no evidence that these techniques can reduce signal loss related to compression and CA blur. Additional research is needed to determine the compounding effect of LLE and SIC on TACSI performance. A demonstration of these techniques is outside of the scope of this study, since their implementation would require non-trivial system modifications.

This technique has broad reaching implications for ultra-fast photography, as it makes the acquisition of dynamic information from both intrinsically and externally sustained radiant entities possible. Intrinsically radiant entities emit photons in response to underlying chemical and physical interactions, which could include thermal radiation\cite{Basu2009}, bioluminescence\cite{Badr2011}, radioluminescence\cite{Klein2019}, plasma discharge\cite{Conrads2000}, and sonoluminescence\cite{Margulis2002}. Externally sustained radiant entities must reflect or scatter photons from an independent light source, or re-emit absorbed photons by fluorescence or phosphorescence. This category is exceptionally broad, as it includes photography at all scales.

\subsection{Limitations} \label{sec:discussion:limitations}

The demonstrated 30 µs temporal resolution is intrinsically linked to the 500 $\mu s$ streak duration used in this study. With the selected lens configuration, the galvo-scanners oscillated at 66 Hz. The specified maximum continuous scan frequency is 250 Hz, which would produce streak durations near 100 $\mu s$, with 5 $\mu s$ temporal resolution. The maximum frame rates that can be achieved with galvo-scanners are around 1-2 MFPS\cite{Liu2019a}, more than sufficient for the vast majority of biological processes. 

The Obis laser used to collect microbead images had a digital trigger, but could not provide sufficient irradiance for CHO-K1 cells. The Genesis laser used to collect CHO-K1 cell images was limited to a $\sim$1 ms pulse duration by a mechanical shutter. This could explain why microbead videos reconstructed, albeit with severely degraded features. A 1 ms pulse duration implies that the single-axis compression ratio was at least 87, above the threshold suggested by Ma et al.\cite{Ma2021a}. This could have contributed to the failed cell video reconstructions. While a chopper wheel could have solved this issue, this demonstrates the ability of TACSI to control the pulse duration using the second axis. Even if the pulse duration was reduced to 500 $\mu s$, the single-axis compression ratio would still have been $\sim$87 because the interval over which overlap occurs was 150 $\mu s$ based on the diameter of the cell. This remains 10-fold higher than the TACSI compression ratio using the same parameters.

Overcoming the camera's rolling shutter effects required long exposure durations. The object under investigation was continuously illuminated for a duration no longer than half of the galvo-scanner's oscillatory period to avoid integrating photons from multiple passes. Using a camera with a global shutter would eliminate the need to shutter the source, but generally results in higher read noise.

The TACSI method can be generalized to other streak mechanisms if higher frame rates are necessary, including polygon scanning mirrors and potentially a streak camera. A streak camera based TACSI system would require placing a CA at the camera plane after the streak tube. This is required to ensure the object streak rate is in excess of the CA streak rate. The decay time of the phosphor may result in streak rate or application limitations\cite{Davis1992}. 

Placing a galvo-scanning mirror at the Fourier plane of a telescope creates an aperture that can reduce throughput. Larger mirrors will have lower oscillatory rates due to inertia. Faster streak rates can be obtained at the cost of lower throughput by increasing the distance from the streak shearing mirror to the imaging lens, proximal to the camera. Higher resolution video could be obtained by using higher numerical aperture optics in the streak shearing relay to image smaller CA elements, limited by the camera's pixel pitch.

\subsection{Future work} \label{sec:discussion:future_work}

This work provides a new mechanism to investigate the theoretical limits of compressed streak modalities. The hypothesis investigated herein focused on the potential for an object with well defined boundaries to create a flash and shutter effect. However, preliminary work investigating two-axis simulations generated using 31 scenes from the Cave Hyperspectral Dataset, demonstrated in SI Appendix, Figure \ref{supp:fig:cave_simulations} and \ref{supp:fig:face_detection}, suggests that this technique can be applied more generally. An example of the simulated single-axis and TACSI compressed streak images can also be observed in SI Appendix, Figure \ref{supp:fig:jellybean_dataset}, along with the CA and a ground truth frame. A full investigation into the extent of this generalization is outside of the scope of this manuscript. Supplemental Video 5 and 6 show single- and two-axis reconstructions of the face\_ms scene from the Cave multi-spectral dataset\cite{Yasuma2010}.

It should be emphasized that TACSI does not directly influence the spatial resolution of the acquired video. The improvement in imaging fidelity results from reconstruction artifact reduction. The spatial resolution of a compressed streak imaging system is controlled by the diffraction limit of the optics, pixel size, and compressed sampling. By switching to a 1.42 NA 60x oil immersion objective instead of the 1.2 NA 60x water immersion objective used in this study, the maximum spatial resolution could be improved by $\sim$18\%. It is also possible to use a camera with smaller pixel size. For example, a Hamamatsu Orca-Flash 4.0 v3 with 1x1 binning, has 6.5 $\mu m$ pixels. The Tucsen Dhyana 95 used in this study has 11 $\mu m$ pixels and was constrained to 2x2 binning due to an inability to resolve smaller CA elements with the chosen optics. This limited the effective pixel size to 22 $\mu m$. Changing the camera could improve the spatial resolution by over 3-fold. Reducing the binning factor and minimizing the pixel size requires a high numerical aperture collection optic for the CA. The 100 mm (AC508-100-A-ML, Thorlabs) 2” achromatic lens used in this study has an estimated 0.25 NA. It is possible to use a 2x stereoscopic objective\cite{Liang2017} (MVPLAPO 2XC, Olympus) with a 0.5 NA to further improve the system resolution. The galvo-mirrors were 5 mm leaflets, which may have restricted the aperture at the Fourier plane. If so, this could produce a lowpass filter effect, which can be mitigated with a larger galvo-mirror. The inertia of the galvo-mirror should be considered when using a larger mirror as this could limit the maximum oscillatory frequency.

TACSI may also be useful for optimization of hyperspectral imaging via coded aperture snapshot spectral imaging (CASSI)\cite{Wagadarikar2008}, which shares several key attributes with CSI. CASSI has recently demonstrated applications in fundus imaging that may benefit from two-axis compressed imaging\cite{Zhao2023}.

Compressed sampling theory requires sparsity and incoherent sampling. Sparsity is typically assumed in most imaging applications. In CSI the CA is designed to capture each time step of the scene incoherently. However, streaking produces some overlap between the adjacent time steps. When imaging an object that moves with respect to the CA, the pattern coding the object's position varies with each time step. A stationary object, by comparison, will be encoded by an invariant pattern, which could exacerbate the degree of overlap between mask elements. It is possible to use neural network based optimization strategies to both optimize the encoding pattern\cite{Wang2019} and improve video reconstructions\cite{Liu2021,He2024,Yang2021,Marquez2022}. The effect that encoding pattern has on reconstruction artifacts at various streak angles was not evaluated.

While an investigation into the effects that the choice of reconstruction algorithm on TACSI was outside of the scope of this work, it remains necessary to determine why the decreased TACSI compression ratio results in an increased BUR when signals are reconstructed using ADMM-PnP. The advantage demonstrated with TACSI does not appear to be algorithm dependent. The relative intensity changes seen in Figure \ref{fig:temporal_res_and_sensitivity}A and \ref{fig:temporal_res_and_sensitivity}B, which were derived from compressed streak images reconstructed with TwIST.

\section{Materials and methods}\label{sec:materials_and_methods}

\subsection{Optical system}
\label{sec:methods:optical_system}

\begin{figure}[ht!]
\centering
\includegraphics[width=0.9\textwidth]{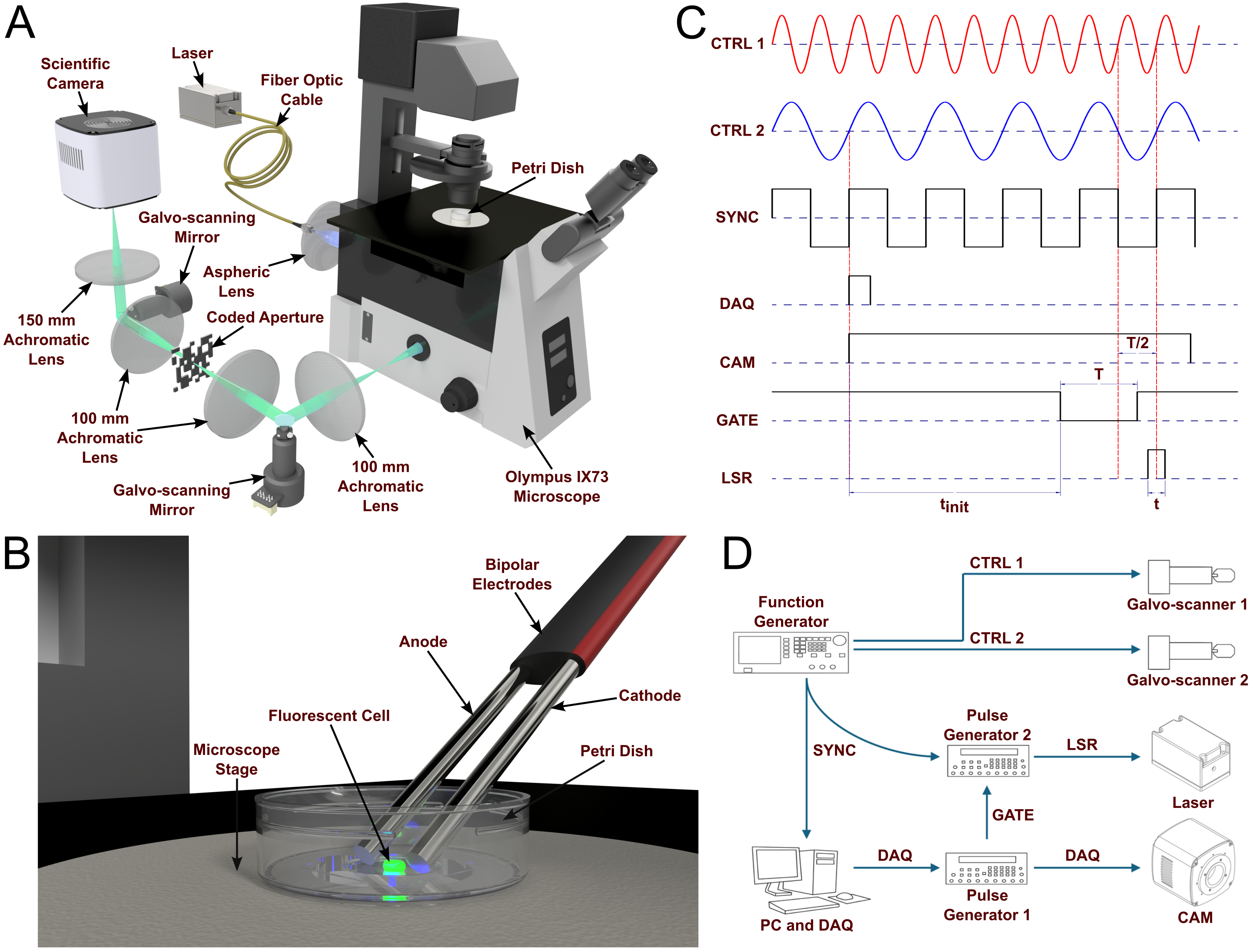}
\caption{Optical system, timing, hardware, and control signals. A CAD model of the TACSI optical system can be seen in sub-panel (A). The placement of the bipolar electrodes in relation to the fluorescent cell can be observed in sub-panel (B). The cell and electrodes are not drawn to scale. Sub-panel (C) shows the control signals and relative timing, and sub-panel (D) displays the control hardware and signal pathways.}
\label{fig:optical_diagram}
\end{figure}

A TACSI system was integrated into the side-port of a commercial inverted widefield microscope (IX73, Olympus). The configuration of this optical system is depicted in Figure \ref{fig:optical_diagram}A. The IX73 consisted of a 60x water immersion objective lens with a numerical aperture of 1.2 (UPLSAPO60XW, Olympus), 488 nm dichroic mirror (FL-007080, IDEX Health \& Science), 488 nm long pass filter (FL-008552, IDEX Health \& Science), and a 180 mm tube lens. Flat field illumination was achieved by imaging the 1000 $\mu$m core of a multimode fiber optic cable to the microscope's sample plane with an aspheric collimation lens. Images of CHO-K1 cells were acquired with a 488 nm laser (Genesis CX488-3000 STM, Coherent) at $\sim$7 $kW/cm^{2}$ irradiance, while fluorescent beads (Focal check Test Slide \#1, Invitrogen) were obtained between roughly 0.1 and 0.4 $kW/cm^{2}$ illumination at 488 nm using a laser (Obis LX, Coherent) capable of analog and digital modulation. A chrome on quartz CA (CMTAMMK210816, PhotomaskPORTAL) was positioned at the image plane between the two streak telescopes. A pair of 2" achromatic lenses with 100 mm focal lengths (AC508-100-A-ML, Thorlabs) resulted in unity magnification between the microscope and the CA. The second streak telescope had a 100 mm achromat (AC508-100-A-ML, Thorlabs) and a 150 mm achromat (AC508-150-A-ML, Thorlabs), resulting in a 1.5x magnification between the CA and the camera. The image was relayed to a scientific complementary metal-oxide semiconductor (sCMOS) camera (Dhyana 95 v1, Tucsen). The pulse duration of the Genesis was controlled by a mechanical shutter and controller (T132, Uniblitz), which limited the minimum duration to around 1 ms.

A shearing interferometer was used to position the galvo-scanners (GVS002, Thorlabs) at the Fourier plane of each 4f telescope. The procedure is as follows. A collimated laser beam was aligned using a pair of irises. The galvo-scanner was rotated until the beam was retro-reflected back through both irises. The achromatic lens was then positioned such that the retro-reflected beam was collimated. At each step, the collimation was evaluated with a shearing interferometer. This procedure was repeated for the second achromatic lens in the telescope. The galvo-scanner was then rotated to 45$^{\circ}$. The collimated beam was directed through the telescope, and the shearing interferometer was used to verify that the beam remained collimated as it exited the telescope.

\subsection{Streak timing}
\label{sec:methods:streak_timing}

Figure \ref{fig:optical_diagram}C and \ref{fig:optical_diagram}D show the TACSI system control signals and timing hardware, respectively. Two galvo-scanners were rastered continuously with sinusoidal input signals (CTRL1 and CTRL2) generated by a 2-channel function generator (DG1022Z, Rigol). Galvo-scanner 1 (CTRL1) controlled the position of the sample with respect to the CA image plane, while galvo-scanner 2 (CTRL2) controlled the position of the CA with respect to the camera sensor plane. It is important to note that in Figure \ref{fig:optical_diagram}C, CTRL1 is depicted as having twice the frequency of CTRL2. However, any integer ratio of control frequencies is acceptable. Non-integer ratios are also possible but will require a phase offset. All events were synchronized to the galvo-scanner using a square wave (SYNC) referenced to CTRL2. The galvo-scanner control signals were driven with an integer frequency ratio.

Upon user interaction with custom LabView software, a data acquisition card (PCIe-6321, National Instruments) counter was initialized, such that a 100 $\mu$s trigger pulse (DAQ) was instigated by the next rising edge of SYNC. The DAQ pulse triggered a digital delay generator (Pulse Generator 1, DG645, Stanford Research Systems), which relayed the DAQ pulse to trigger the sCMOS (CAM). The camera was continuously integrated over a 160 ms acquisition time to overcome rolling shutter effects, and required a minimum 26 ms delay $t_{init}$ prior to image acquisition. This delay was mediated by a gating signal (GATE) delivered to the inhibit port on a second digital delay generator (Pulse Generator 2). When the GATE signal was pulled low, Pulse Generator 2 was triggered by the next falling edge of the SYNC signal. The laser was then initiated after a $(T-t)/2$ delay, where $T$ is the period of CTRL2 and $t$ is the sample streak duration as defined by Equation (\ref{eq:streak_duration}):
\begin{equation}
\label{eq:streak_duration}
t=\frac{1}{\pi f}sin^{-1}\bigg(\frac{V_{H}-V_{L}}{2A}\bigg) \,.
\end{equation}
The laser was digitally triggered to prevent continuous exposure due to the continuously rastering galvo-scanners. The sample streak duration is controlled by the frequency $f$ and amplitude $A$ of CTRL1. Both CTRL1 and CTRL2 were generated with a 10 $V$ amplitude to maximize streak linearity. The high and low threshold voltages, $V_{H}$ and $V_{L}$, are the voltages at which the center line of the sample plane intersects with the left and and right camera sensor edges, respectively.

\subsection{System characterization}
\label{sec:methods:characterization}

The sensitivity and temporal resolution for the single-axis and TACSI methods were evaluated in Figure \ref{fig:temporal_res_and_sensitivity}. The temporal resolution of the single-axis and TACSI systems was determined by calculating the intensity contrast from fluorescent microbeads modulated with a sinusoidal laser intensity (peak-to-trough). Compressed streak images were acquired for modulation frequencies ranging from 10 $kHz$ to 45 $kHz$ ($N=3$). The temporal resolution was defined as the period of the modulation signal that resulted in a 50\% contrast. Contrast $C$ was fit with a lowpass filter function of the form:
\begin{equation}
\label{eq:lowpass_fit}
C(f)=\frac{A}{1+\big(\frac{f}{f_c}\big)^n} \,,
\end{equation}
where $A$ is the signal amplitude, $f$ is the laser modulation frequency, $f_c$ is the cut-off frequency, and $n$ is the roll-off exponent. The goodness of fit was evaluated using the $R^2$ and normalized root mean square error (NRMSE).

The sensitivity for the single-axis and TACSI system was defined by the slope of the calculated contrast (peak-to-trough) vs the relative change in the laser intensity at a constant sinusoidal modulation frequency of 10 kHz. Compressed streak images were acquired for relative changes in the modulation signal from 10\% to 100\% ($N=3$). Linear fits were evaluated using the $R^2$ and NMRSE. To calculate the relative change in counts for the sensitivity data, each profile was fit to a polynomial of order 4 to estimate the background intensity. The relative difference from the background intensity was calculated at each time point. The resulting difference profiles were then averaged.

\subsection{Fluorescent microbead imaging}
\label{sec:methods:fluorescent_microbead_imaging}

The single- and two-axis empirically measured fluorescent microbead streak images shown in Figures \ref{fig:fundamental_concept}K and \ref{fig:fundamental_concept}L were acquired with a 500 $\mu s$ streak duration. The horizontal object speed and the vertical streak speed were roughly 45 m/s and 4.5 m/s at the camera sensor, respectively. A 10 kHz sinusoidal intensity modulation was produced using an Obis LX laser with analog inputs between $\sim$1.4 and $\sim$3.4 V. The analog voltage varied due to the fluorescence efficiency of the bead under investigation and the CA position. The laser power was adjusted for each measurement to produce roughly 5,000 and 12,000 counts in 50x50 ROIs placed within the streak path and centered on a local minima and maxima, respectively. ROIs were selected using an ImageJ macro. This resulted in single-axis streak images with a maximum intensity of around 60,000 counts. Three beads were measured across 3 replicate groups. The CA was repositioned for each replicate. The system response was subtracted using a second order polynomial fit before plotting the profile in Figure \ref{fig:fundamental_concept}I.

\subsection{Cell culture}
\label{sec:methods:cell_culture}

CHO-K1 cells (ATCC CCL-61) were cultured according to standard ATCC protocols. The CHO-K1 cells were propagated in a complete growth medium consisting of Kaighn’s Modification of Ham’s F-12 Medium, 10\% fetal bovine serum, 2 mM L-glutamine, and 1\% by volume 100 U/mL penicillin/streptomycin (ATCC 30-2300). The cells were maintained in an incubator held at 37\textdegree C, 95\% humidity, and 5\% CO\textsubscript{2} in air. Before imaging experiments, cells were passaged, diluted with fresh media, and allowed to settle onto a glass bottom dish (FluoroDish, poly-D-lysine coated, FD35PDL, World Precision Instruments, Inc.).

For membrane potential imaging experiments, the cells were loaded with a solution made using the FluoVolt\textsuperscript{TM} Membrane Potential Kit (F10488, Invitrogen, Thermo Fisher Scientific). Per the manufacturer's protocol, 1 $\mu$L of the FluoVolt\textsuperscript{TM} dye and 10 $\mu$L of the 100X PowerLoad\textsuperscript{TM} concentrate were added to 2 mL of live cell imaging solution (LCIS) to formulate the loading solution. For loading the dye, the cell dishes had their growth media removed, were rinsed with LCIS, and had their media replaced with the loading solution. After loading the solution into the cell dish, the cells were placed into the incubator for 30 minutes for dye-loading. The dish was then rinsed, replaced with 2 mL of LCIS, and brought to the optical system for experiments.

\subsection{Electric pulse delivery}
\label{sec:methods:pulse_delivery}

Microsecond electric pulses were delivered via a pair of platinum iridium bipolar electrodes with a 125 $\mu$m diameter. Electrode spacing was measured prior to data collection at 117$\pm$1 $\mu m$ (inner edge-to-edge). The electrode positioning and spacing was measured prior to data collection using a micromanipulator (Trio\textsuperscript{TM}/MP-245, Sutter Instruments). The electrodes were raised 50 $\mu$m above the petri dish bottom and angled 30\textdegree\ with respect to the plane of the cover slip. Square pulses were generated with a 50 $\mu$s pulse width at 15 $V$ using a general purpose pulse generator (AV-1015-B, Avtech). The electric field strength at the cells, when centered between the electrodes, was predicted to be roughly 800 $V/cm$ by finite-difference time-domain (FDTD) modeling\cite{Ibey2010}. Several conventional streak images were acquired with a closed slit to verify that the pulse timing matched the oscilloscope traces and to ensure that the electric pulse strength did not cause cell movement before acquiring the first compressed streak image. An example of a closed slit membrane potential measurement is shown in SI Appendix, Figure \ref{supp:fig:slit_streak}. The maximum field strengths used here are very close to those used by Kiester et al.\cite{Kiester2021}

\subsection{Image Reconstruction}
\label{sec:methods:reconstructions}

Prior to reconstruction, background was subtracted from the streak images using the average of 30 dark frames. Streak image and coded aperture values were converted to 32-bit floating point and min/max normalized between 0 and 1. The CA images were then converted to binary masks with an intensity threshold. Simulated and measured fluorescent microbead videos were recovered using a compressed streak reconstruction algorithm implemented in MATLAB, employing the plug-and-play alternating-direction method of multipliers (ADMM-PnP) as described by Lai et al.\cite{Lai2020}, which incorporates BM3D denoising\cite{Dabov2009}. The reconstructions were performed with penalty weights set to $\mu_{1} = \mu_{2} = \mu_{3} = 1$ and a noise suppression factor of $\sigma = 20$. The number of iterations was fixed at 100, with a termination ratio of 0.015 and a plateau tolerance of 5. Cellular reconstructions were computed using a GPU accelerated version of TwIST\cite{Ma2021a} with total variation regularization. For both two-axis and single-axis video reconstructions, the image frames were registered along the CA streak axis. Additionally, the two-axis streak frames were translated along the object streak axis to nullify the induced movement after reconstruction.

\subsection{Streak Simulations}
\label{sec:methods:simulations}

Simulations were developed with custom Python code to assess the reconstruction advantage of TACSI over conventional single-axis reconstructions in the absence of noise sources. High resolution videos were generated with a frame rate of 7.168 MFPS with a sequence depth of 3885 frames. Image frames spanned 1020 rows and 1024 columns with a pixel pitch of 22 $\mu$m. Images of intensity-modulated 10 $\mu$m fluorescent microbeads were simulated using OpenCV. The intensity was adjusted with a 10 kHz sinusoidal function. Single-axis compressed streak images were generated from video of a stationary bead, while two-axis streak images were generated with the bead moving at 45.056 m/s horizontally across the image frame. Binary pseudo-random CAs were generated with 1020 rows and 1024 columns. Each binary CA element spanned 3x3 pixels with the value of each 3x3 group drawn from a Gaussian distribution. Streak images were generated by affine transforming each frame with an order of magnitude slower CA streak rate orthogonal to the path of the bead. An integer ratio was chosen to accomodate the sinusoidal galvo-scanner control signals when acquiring laboratory data. All calculations were performed on 32-bit floating point values, then converted to 16-bit TIF to approximate laboratory acquired images.

\subsection*{Abbreviations}
\begin{tabbing}
    \hspace{2.5cm} \= \kill 
    TACSI \> Two-axis compressed streak imaging \\
    CSI \> Compressed streak imaging \\
    UHS \> Ultra-high-speed \\
    CUP \> Compressed ultra-fast photography \\
    sCMOS \> Scientific complementary metal-oxide semiconductor \\
    EMCCD \> Electron multiplying charge-coupled device \\
    CA \> Coded aperture \\
    PEF \> Pulsed electric field \\
    DMD \> Digital micromirror device \\
    LC-SLM \> Liquid crystal spatial light modulator \\
    CS \> Compressed sensing \\
    BUR \> Bit utilization ratio \\
    GFP \> Green fluorescent protein \\
    CASSI \> Coded aperture snapshot spectral imaging \\
    LCIS \> Live cell imaging solution \\
    FDTD \> Finite-difference time-domain \\
    ADMM-PnP \> Plug-and-play alternating direction method of multipliers \\
\end{tabbing}

\backmatter

\bmhead{Acknowledgments} Work contributed by SAIC was performed under United States Air Force Contract No. FA8650-19-C-6024. J.N.B. received funding from the Air Force Office of Scientific Research under award 23RHCOR002. X.L. and J.L. received funding from the Natural Sciences and Engineering Research Council of Canada (RGPIN-2024-05551). V.V.Y. received partial funding from the Air Force Office of Scientific Research (FA9550-20-1-0366, FA9550-20-1-0367, FA9550-23-1-0599), the National Institutes of Health (NIH) (R01GM127696, R01GM152633, R21GM142107, and 1R21CA269099), and is supported by NASA, BARDA, NIH, and USFDA, under Contract/Agreement No. 80ARC023CA002.

\bmhead{Author contribution} M.A.K. conceived the idea, designed the optical system, wrote the scripts to generate the simulations, wrote the Labview code to operate the system, designed the experiments, reconstructed all videos, formulated the equation for the compression ratio, created all figures, analyzed the results, and wrote the manuscript. M.A.K. and S.P.O. assembled the hardware, aligned the optical system, and acquired all experimental results. S.P.O. cultured the cells and loaded the voltage sensitive dye. All authors participated in the discussion, analysis, and editing of the manuscript. J.L. and X.L. provided additional assistance with video reconstructions. J.N.B. and V.V.Y. supervised the project.

\bmhead{Data availability} The datasets used and/or analysed during the current study are available from the corresponding author on reasonable request.

\bmhead{Conflict of interest} The authors declare that there are no financial interests, commercial affiliations, or other potential conflicts of interest that could have influenced the objectivity of this research or the writing of this paper.

\begin{appendices}

\section{Streaked illumination}
\label{supp:sec:streaked_illumination}

\begin{figure}[ht!]
\centering\includegraphics[width=13cm]{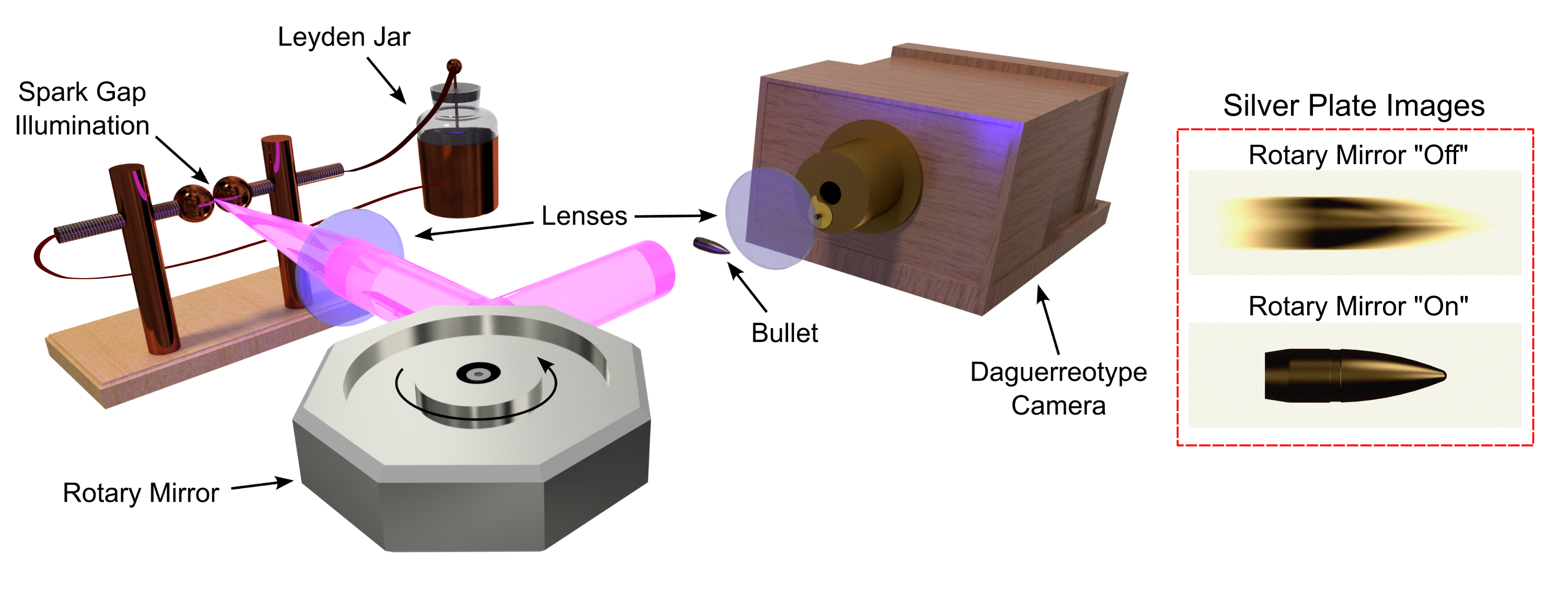}
\caption{Artistic rendering of the 1893 bullet imaging experimental setup. The Leyden jar provided the voltage necessary to generate a flash at the spark gap electrodes. The rotating mirror streaked the flash across the bullet, resulting in a substantially shorter exposure duration. Simulated daguerreotype silver plate images show the differences between images with the rotary mirror on or off.}
\label{supp:fig:streaked_source_1893}
\end{figure}

The fundamental concept used in two-axis compressed streak imaging (TACSI) to reduce motion blur was derived from a streaked illumination method pioneered in 1893 by Boys et al. developed for high speed bullet imaging\cite{Boys1893}. Figure \ref{supp:fig:streaked_source_1893} shows an artistic rendering of the key components from the landmark 1893 experimental setup. The wires were not drawn to simplify the diagram. In the landmark experiment, a flash was generated when the bullet completed a circuit between the Leyden jar and a spark gap electrode. By translating the illumination source, the exposure duration can be shortened substantially.
In TACSI, the bullet is analogous to the translating image of the coded aperture (CA). By translating a continuously illuminated object, the elements of the mask are exposed for a shorter duration than would be possible without translation.

\pagebreak

\section{Compressed streak imaging microbead simulations}
\label{supp:sec:microbead_simulations}

\begin{figure}[ht!]
\centering\includegraphics[scale=0.25]{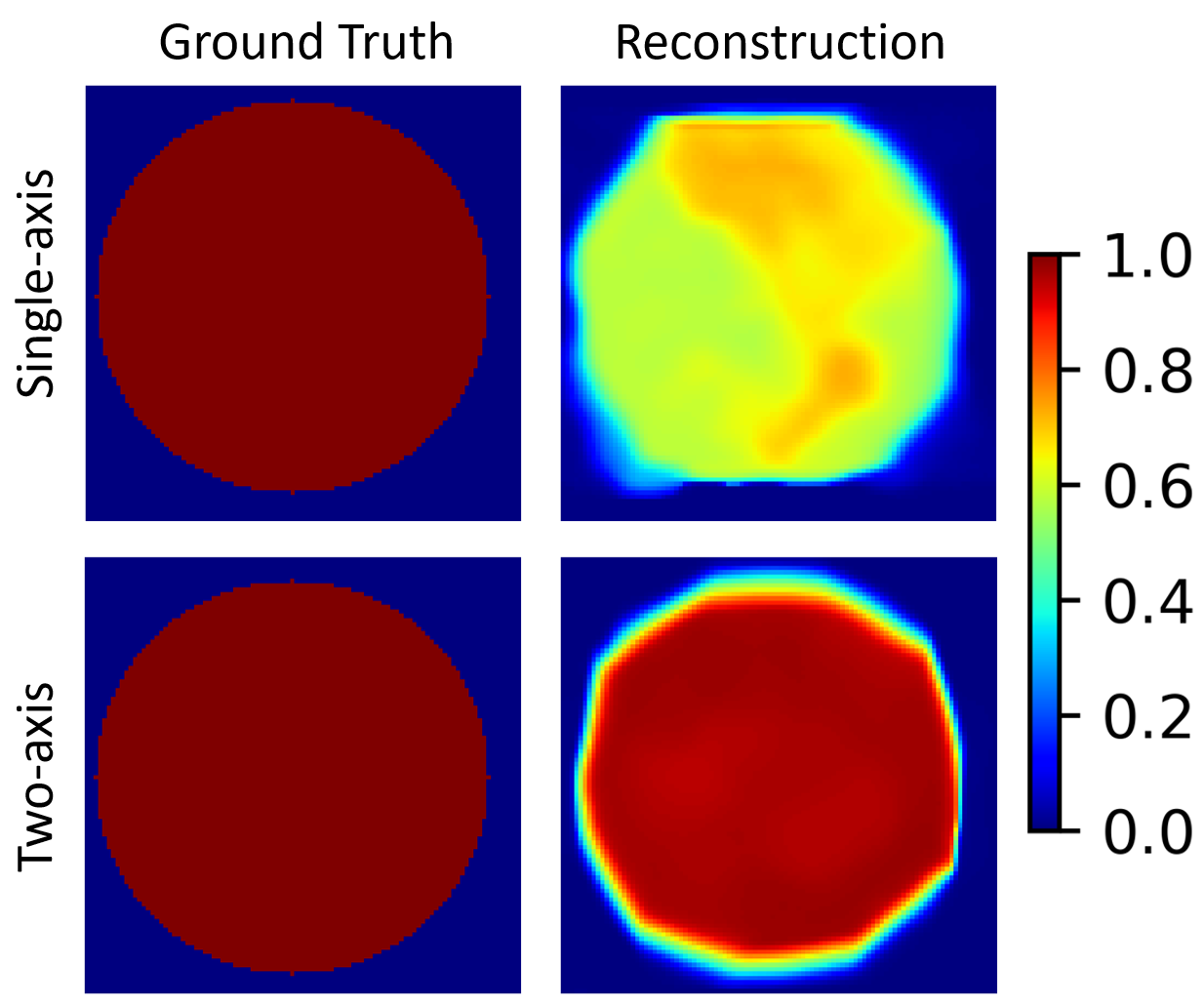}
\caption{Comparison between normalized frames from reconstructed single- and two-axis videos and their ground truth.}
\label{supp:fig:bead_simulation_comparison}
\end{figure}

Single frames from simulated single- and two-axis reconstructed videos are compared to their ground truth in Figure \ref{supp:fig:bead_simulation_comparison}. Table \ref{supp:tab:microbead_metrics} shows improvements in peak signal to noise ratio (PSNR), structural similarity index measure (SSIM), and the bit-utilization ratio (BUR) derived from reconstructions from simulated microbeads using the plug-and-play alternating direction method of multipliers (ADMM-PnP). This improvement was obtained with the object speed set 10-fold higher than the streak speed. Additionally, the resulting compression ratio was reduced by a factor of 10.

\begin{table}[ht!]
\caption{Comparison between single- and two-axis microbead simulation metrics.} 
\label{supp:tab:microbead_metrics}    
\begin{tabular}{lll} 
\hline\hline
Metric & Single-axis  & Two-axis \\
\hline\hline
PSNR (dB)    & 15.9   & 18.3 \\
SSIM    & 0.524   & 0.740 \\
BUR (\%)    & 3.62   & 15.8 \\
CR    & 50.8  & 5.05 \\
\hline\hline
\end{tabular}
\end{table} 

\pagebreak

\section{Cave dataset compressed hyperspectral simulations}
\label{supp:sec:cave_simulations}

\begin{figure}[ht!]
\centering\includegraphics[width=13cm]{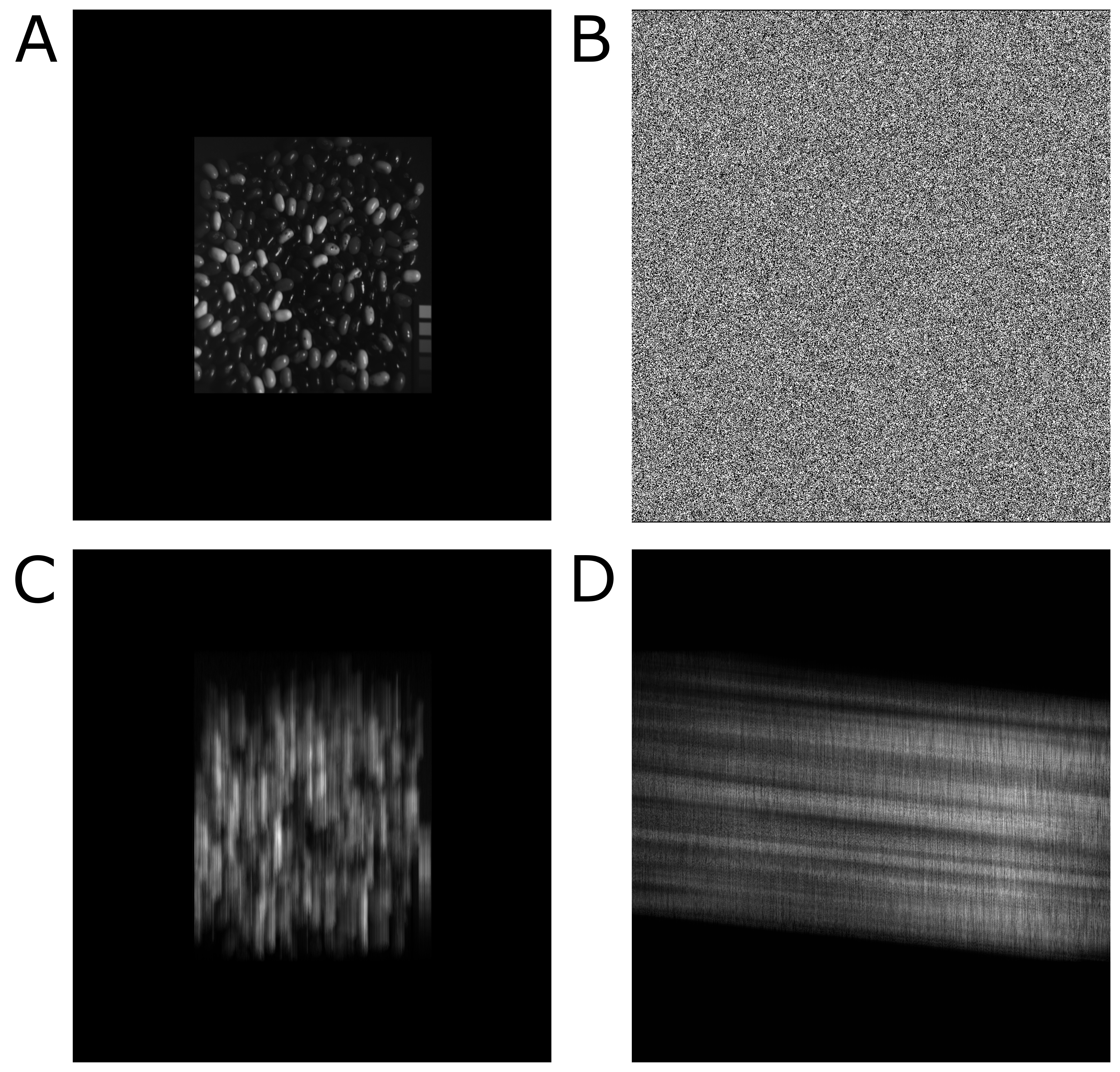}
\caption{Jellybeans streak simulation dataset. (A) A single frame from the ground truth video. (B) The coded aperture used in the single- and two-axis streak procedures. (C) The single-axis compressed streak image. (D) The two-axis compressed streak image.}
\label{supp:fig:jellybean_dataset}
\end{figure}

\begin{figure}[ht!]
\centering\includegraphics[width=13cm]{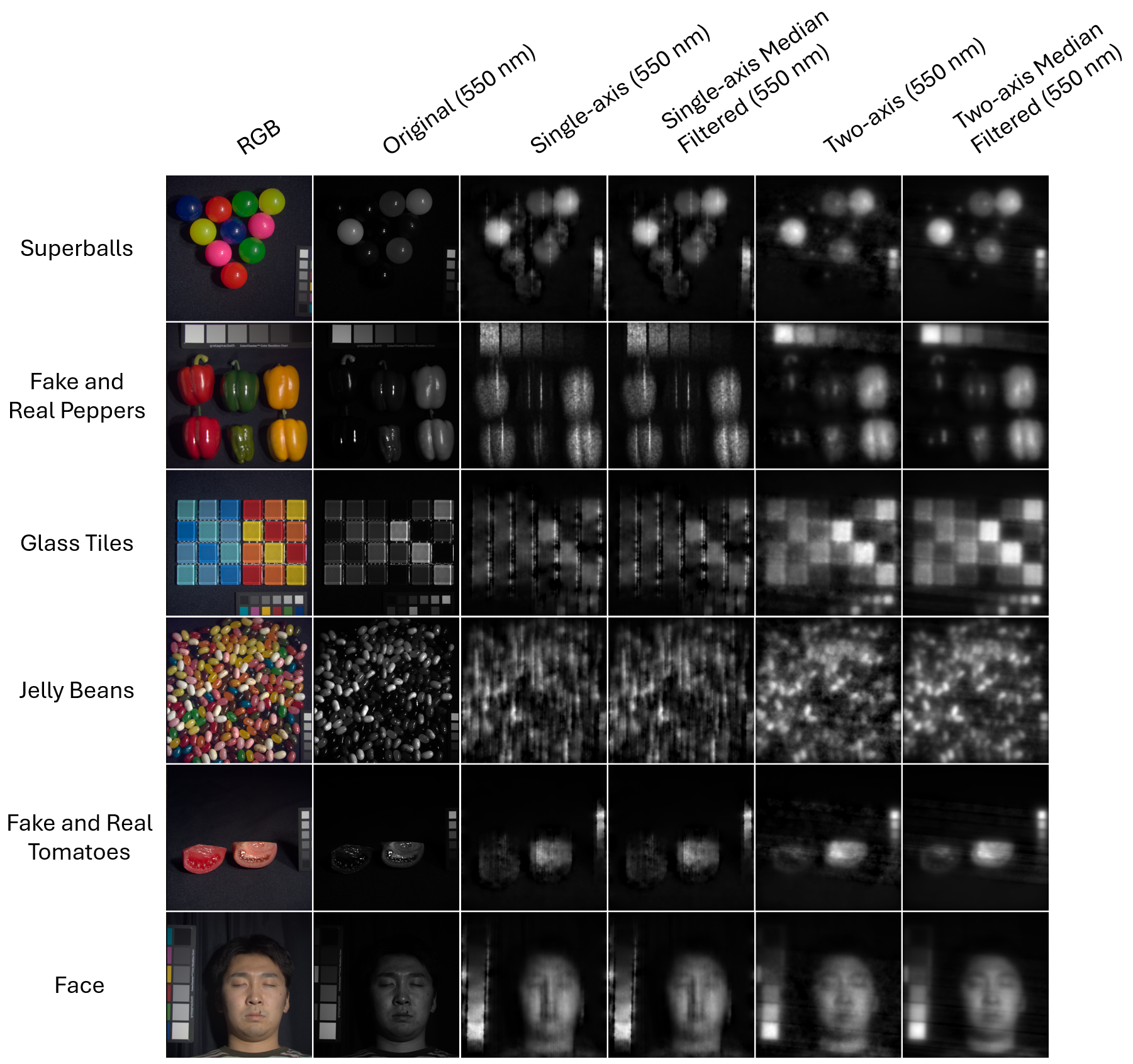}
\caption{Compressed hyperspectral simulations using complex scenes. Scenes from the Cave Dataset were used to generate single- and two-axis compressed hyperspectral simulations. The original image at 550 nm (yellow) is compared to the 550 nm frames from the reconstructed single- and two-axis videos. An RGB image is included to assist with color intensity comprehension. The temporal median filtered 550 nm image for each modality shows greater artifact reduction for two-axis images.}
\label{supp:fig:cave_simulations}
\end{figure}

Improvements in reconstructed video quality may extend beyond CSI to compressed hyperspectral imaging (CHI). Simulated streak images were derived from 31 scenes within the Cave Hyperspectral Dataset\cite{Yasuma2010}. An example of the single-axis and TACSI compressed streak images for the "Jelly Beans" CAVE scene is shown in Figure \ref{supp:fig:jellybean_dataset}, along with the CA and a ground truth frame. Comparisons between the original image, single-axis, and two-axis at 550 nm (yellow) are shown in Figure \ref{supp:fig:cave_simulations}. An RGB image is included to assist with interpreting the color intensities.

Two-axis reconstructions require an extra translation step to nullify the induced object movement. This step results in a static object and simultaneously translates the positions of the reconstruction artifacts. Because the artifacts move from frame to frame, they can be removed using a temporal median filter.

\begin{figure}[ht!]
\centering
\includegraphics[width=0.9\textwidth]{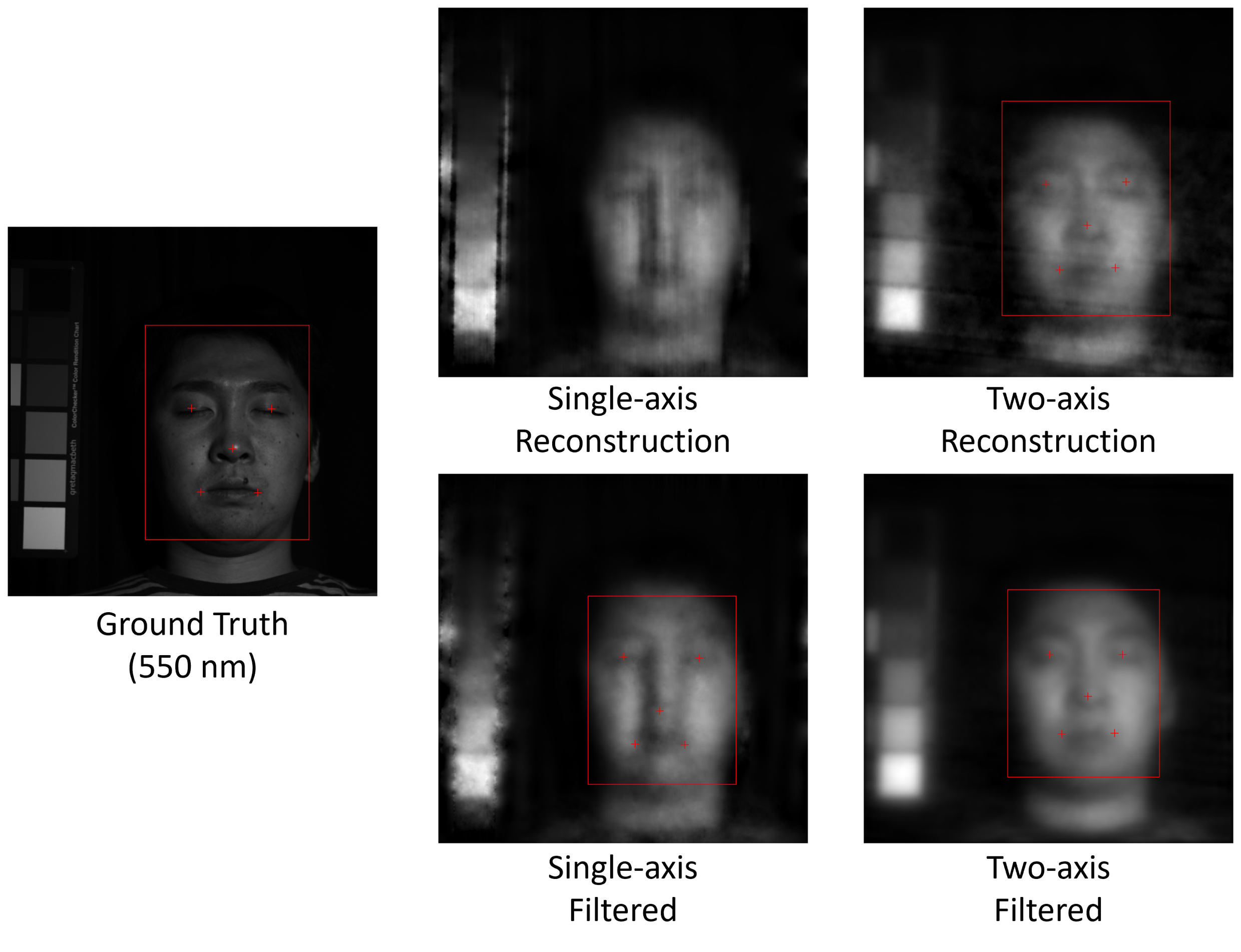}
\caption{Two-axis reconstruction enables face-tracking. The Retina Face face tracking algorithm was applied to the 550 nm ground truth, raw reconstructed single-axis, raw reconstructed two-axis, and temporal median filtered images for each modality with custom masks applied to block diagonal or vertical striations. Red bounding boxes surround the faces, with markers positioned at the eyes, nose, and corners of the mouth, in all images except for the raw single-axis reconstruction.}
\label{supp:fig:face_detection}
\end{figure}

Face detection was performed using the RetinaFace\cite{Deng2019,serengil2020lightface,serengil2024lightface} on the Face scene from the Cave Hyperspectral Dataset. Figure \ref{supp:fig:face_detection} shows the results from applying the face detection network to the ground truth image, single-axis reconstruction, and two-axis reconstruction. The network is unable to detect facial features in the single-axis reconstruction, whereas the two-axis reconstruction accurately predicts the location of the face, eyes, nose, and corners of the mouth. Retina Face was able to detect facial features in the single-axis reconstructions after applying a custom filter to the image FFT. The vertical (or diagonal) lines were masked along with the high frequency features in both reconstructions.

\pagebreak

\section{Closed slit streak imaging}
\label{supp:sec:closed_slit}

\begin{figure}[ht!]
\centering
\includegraphics[width=0.9\textwidth]{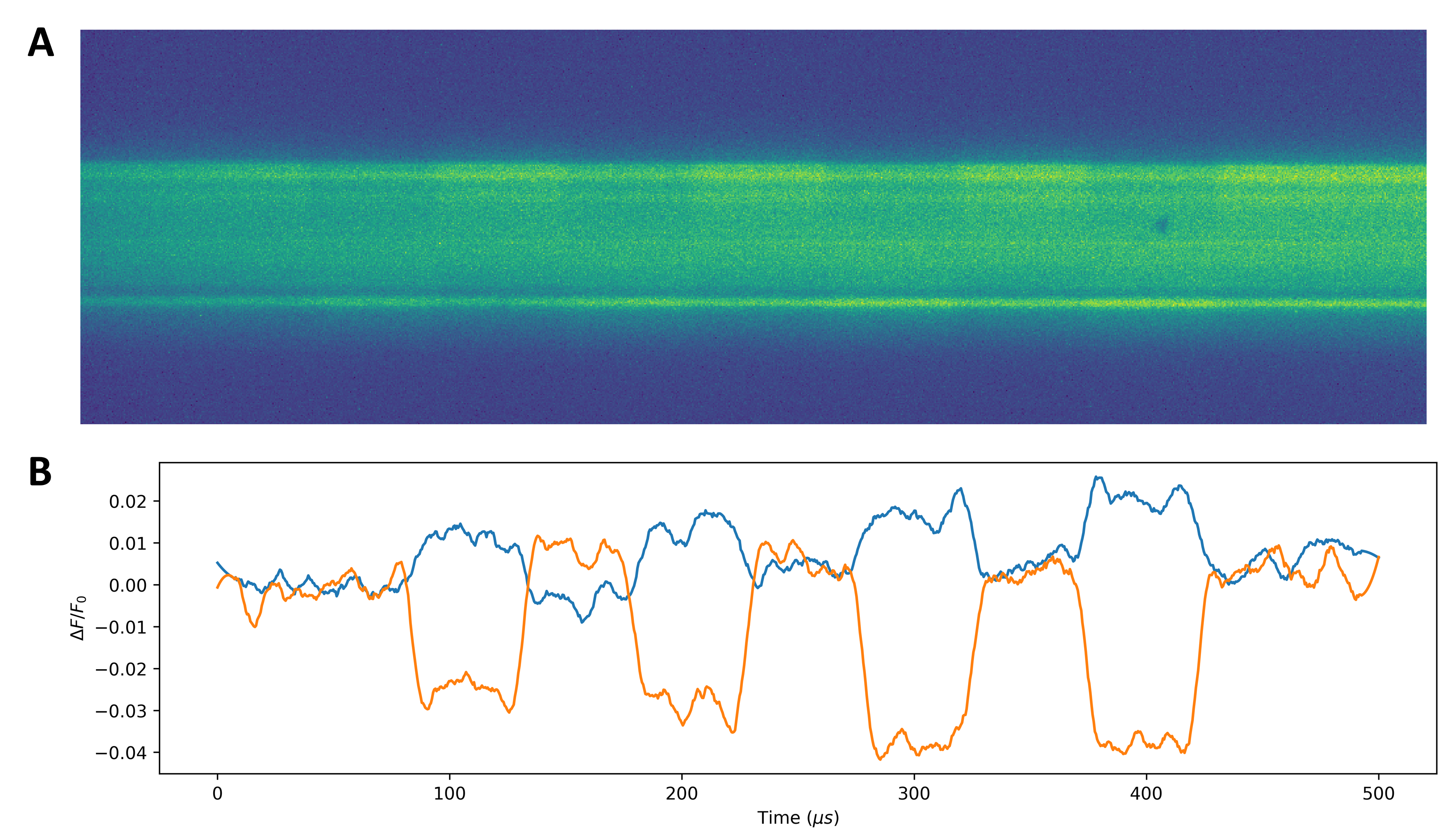}
\caption{Closed slit streak image. Image of a CHO-K1 cell loaded with Fluovolt voltage sensitive dye during electric pulse application with the mechnical slit closed and the coded aperture removed. The intensity at the membrane proximal to the anode increased in intensity, while the membrane proximal to the cathode decreased in intensity. The straight horizontal cell edges demonstrate that the cell did not move during the pulse.}
\label{supp:fig:slit_streak}
\end{figure}

Prior to acquiring CSI images of CHO-K1 cells, electric pulse timing and strength was visualized with a closed slit streak image to ensure that the pulses occurred during the acquisition window and that the pulses would not cause the cells to move. Figure \ref{supp:fig:slit_streak} shows a representative streak profile from a CHO-K1 cell containing Fluovolt voltage sensitive dye. Signal at the anode (blue) increases while signal at the cathode (red) decreases upon pulse delivery. The pulse strength was insufficient to cause the cell to move, as indicated by the straight cell edges.

\section{TACSI Compression Ratio Derivation}
\label{supp:sec:compression_ratio}

Controlling the portion of the camera sensor over which the digital spatiotemporal streak signal is acquired has an effect on the signal's compression ratio, defined generally as in Equation (\ref{supp:eq:general_compression_ratio})\cite{Ma2021a}:
\begin{equation}
\label{supp:eq:general_compression_ratio}
CR = \frac{N_{x}N_{y}N_{t}}{N_{s'}},
\end{equation}
where $N_{x}$, $N_{y}$, and $N_{t}$ are the number of pixels required to convey the spatial information along the x and y coordinate axes and the time information of a full resolution video. $N_{s'}$ refers to the number of spatial pixels after the video has been compressed into a single streak image. Because the spatial information and the temporal information mix, the total number of camera pixel rows $P_y$ must be constrained by:
\begin{equation}
P_y \geq N_{y}+N_{t}-1.
\end{equation}
This implies that the compression ratio definition treats the extent of the signal and not that of the camera sensor.

\begin{figure}[ht!]
\centering
\includegraphics[width=0.9\textwidth]{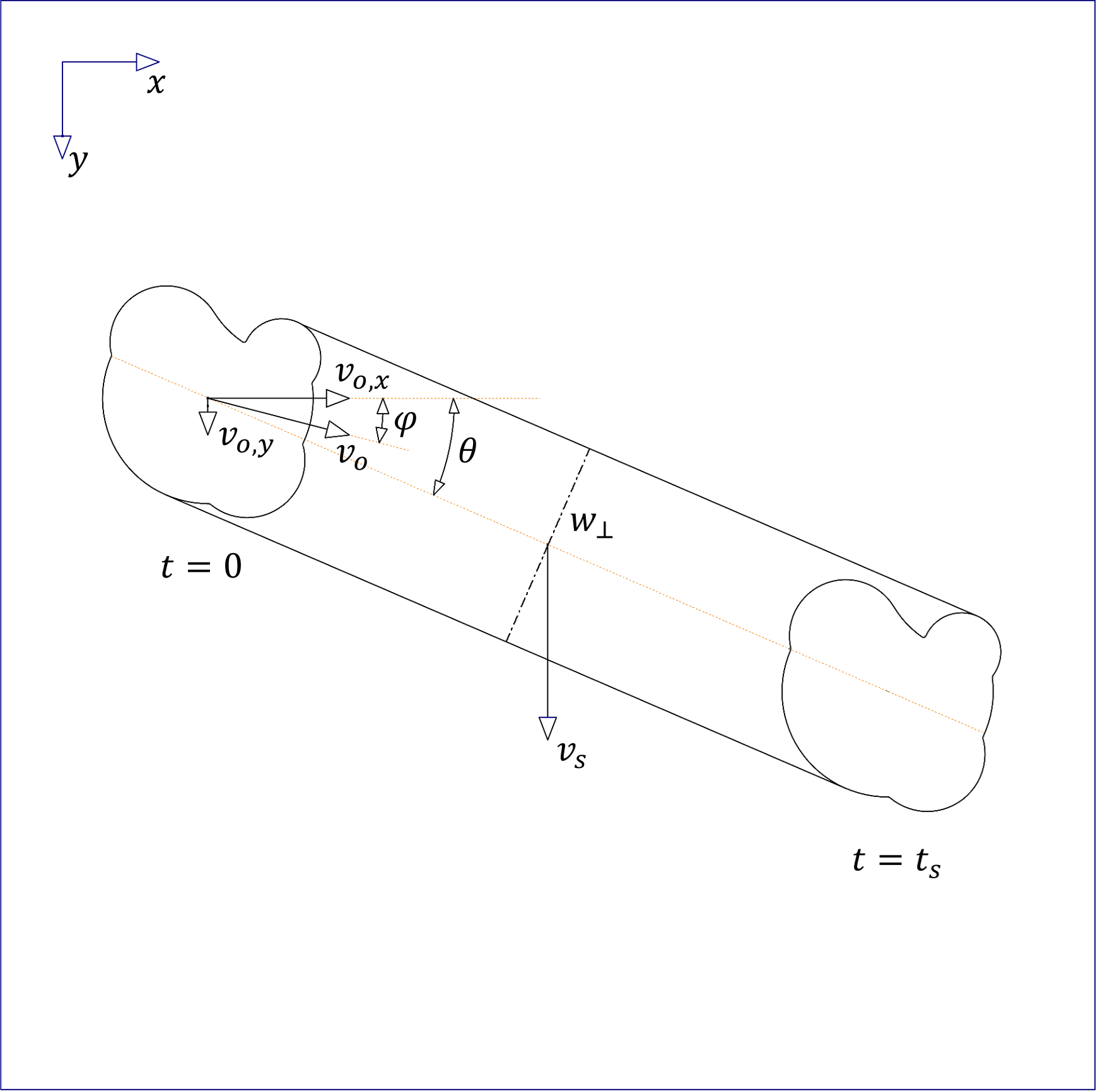}
\caption{Streak imaging conceptual model. The compression ratio can be developed by considering the surface area exposed by a blob with a object velocity $v_{o}$ from time $t=0$ to $t=t_{s}$. The streak velocity $v_{s}$ is the rate at which the image of the CA is propagated with respect to the sensor's vertical axis. The vertical extent of the object along normal to its trajectory is given by $w_{\perp}$. While this model applies to any arbitrary object velocity angle $\varphi$, the TACSI system translates the image of the sample along the horizontal axis of the sensor ($\varphi=0$).}
\label{supp:fig:compression_ratio_model}
\end{figure}

The following is a derivation of the mathematical description of the TACSI compression ratio. Since many scenes are characterized by objects contrasted against dark counts, a general description of the scene should have an amorphous geometry. Using a bounding box surrounding the region of interest may result in an overestimate of the CR. When discussing this topic, the streak velocity $v_{s}$ will be used to refer to the translation rate of the coded aperture image formed at the camera sensor along the sensor's vertical axis. The term object velocity will be used to describe the speed and direction of motion of the image of the object under investigation at the camera sensor plane, with coordinates defined in relation to the horizontal and vertical sensor axes. Figure \ref{supp:fig:compression_ratio_model} depicts the path traced by a blob shaped object defined by the object's velocity and the streak velocity.

The first challenge in describing compression ratio is determining the constraints. The post-reconstruction frame rate can provides a useful constraint for determining the correct number of frames in the high resolution image. This is necessary because the full resolution image is often hypothetical under empirical conditions. The frame rate of a compressed streak imaging system is defined in Equation (\ref{supp:eq:streak_fps}):
\begin{equation}
\label{supp:eq:streak_fps}
FPS=\frac{v_{s}}{p} \,,
\end{equation}
where $v_{s}$ is the streak velocity and $p$ is the pixel pitch of the camera. 

The surface area of the streak in square pixels can be determined from Equation (\ref{supp:eq:streak_surface}):
\begin{equation}
\label{supp:eq:streak_surface}
N_{s'} = \frac{w_{\perp}t\sqrt{v_{o}^{2}+v_{s}^{2}}+s_{o}}{p^{2}}\,,
\end{equation}
where $s_{o}$ is the surface area of the object in square meters, $w_{\perp}$ is the width of the object along an axis perpendicular to its trajectory, and $t$ is the streak duration. The number of image frames in the high resolution video can be found using Equation (\ref{supp:eq:num_frames}):
\begin{equation}
\label{supp:eq:num_frames}
N_{t}=\frac{v_{s}t}{p} \,,
\end{equation}
and the number of spatial pixels as in Equation (\ref{supp:eq:full_surface}):
\begin{equation}
\label{supp:eq:full_surface}
N_{s}=\frac{s_{o}}{p^{2}} \,.
\end{equation}

We can now use Equation (\ref{supp:eq:compression_ratio}) to determine the compression ratio of the streak image:
\begin{equation}
\label{supp:eq:compression_ratio}
CR=\frac{N_{s}N_{t}}{N_{s'}}=\frac{s_{o}v_{s}t}{p(w_{\perp}t\sqrt{v_{o}^{2}+v_{s}^{2}}+s_{o})} \,.
\end{equation}

Let's assume that we want to recover a 200 kFPS video. Then, with our camera's 2x2 binned pixel pitch of 22 $\mu m$, we would need a 4.4 m/s streak velocity along the y-axis based on Equation (\ref{supp:eq:streak_fps}). Further assume that the object is a circular bead with a 2 mm diameter at the camera plane, moving with an object velocity of 44 m/s (10x) along the x-axis. The streak interval is determined by the time required for the bead to travel twice its diameter along the direction of the total velocity, considering both the object and streak velocities. This duration represents the maximum local frame density that arises as the object is convolved with itself across successive time points. Using Figure \ref{supp:fig:compression_ratio_model} as a reference, the overlap time can be generally computed with Equation (\ref{supp:eq:time_for_2diameters_general}):
\begin{equation}
\label{supp:eq:time_for_2diameters_general}
t=\frac{2d_{\theta}}{\sqrt{v_{o,x}^{2}+(v_{o,y}+v_s)^{2}}},
\end{equation}
where $d_{\theta}$ indicates the diameter of the bead taken along the axis parallel to the total velocity vector, with the magnitude of the total velocity indicated in the denominator of Equation (\ref{supp:eq:time_for_2diameters_general}). In the case of our TACSI system, the object velocity travels perpendicularly to the streak velocity ($\phi=0$), allowing for $v_{o,x}=v_o$ and $v_{o,y}=0$. The expression then simplifies to Equation (\ref{supp:eq:time_for_2diameters_TACSI}):
\begin{equation}
\label{supp:eq:time_for_2diameters_TACSI}
t_{\phi=0}=\frac{2d_{\theta}}{\sqrt{v_{o}^{2}+v_s^{2}}}.
\end{equation}
Using Equations (\ref{supp:eq:compression_ratio}) and (\ref{supp:eq:time_for_2diameters_TACSI}) along with the bead geometry, pixel pitch, object velocity, and streak velocity specified earlier in this example, the compression ratio for the object moving horixontally at 44 m/s is approximately 5.1. If we then set the object velocity to 0 m/s, the compression ratio increases to 51.3. This indicates a remarkable 10-fold improvement in the compression ratio in the case of the moving object compared to the stationary object.

\begin{figure}[ht!]
\centering\includegraphics[width=13cm]{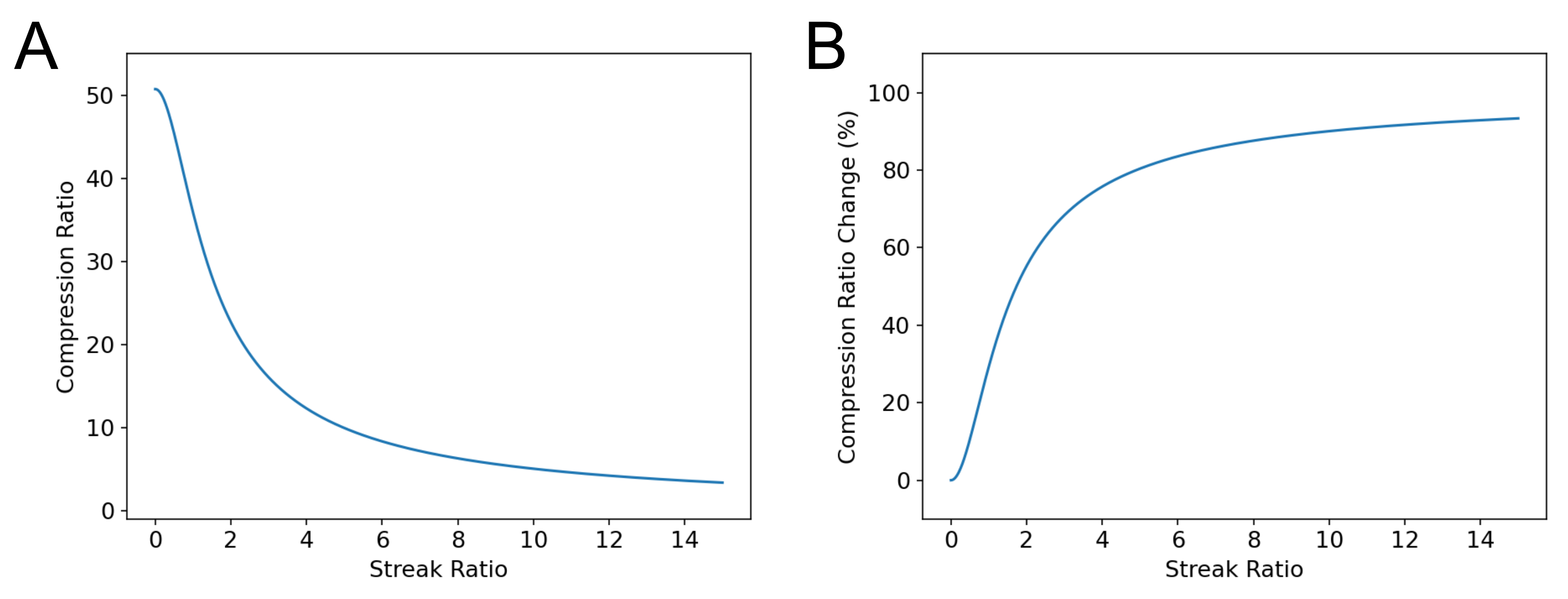}
\caption{Relationship between the compression ratio and the streak ratio. Compression ratio trends can be observed for a circular object with a 2 mm diameter at the camera sensor plane. (A) shows a decreasing compression ratio as the ratio between the object speed and the streak speed increases. (B) shows the compression improvement as a percent with respect to the maximum.}
\label{supp:fig:compression_ratio_plots}
\end{figure}

Figure \ref{supp:fig:compression_ratio_plots} shows the compression ratio as a function of the ratio between the object speed and the streak speed. These results assume that the camera sensor is always wide enough to encode the entire streak interval. A streak ratio of 0 implies the object is stationary, while a ratio 10 would correspond to a 44 m/s horizontal object velocity.

\pagebreak

\section{Discretized TACSI Compression Ratio}
\label{supp:sec:discrete_compression_ratio}

\begin{figure}[ht!]
\centering\includegraphics[width=13cm]{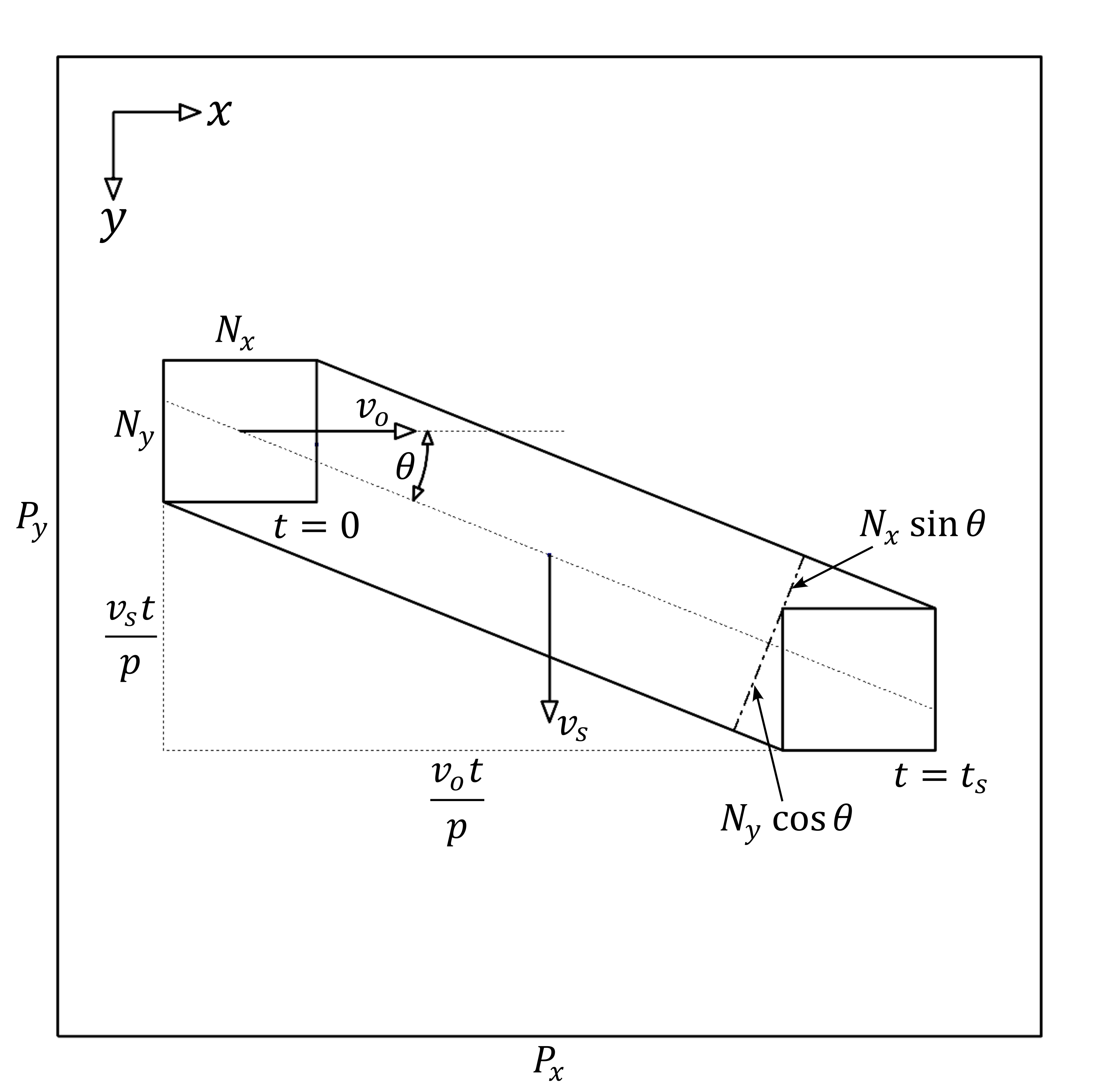}
\caption{Streak imaging conceptual model for a rectangular scene. A discrete compression ratio model for a scene with a rectangular scene geometry can be developed by considering the model depicted above. The box that defines the perimeter of the model represents the imaging sensor.}
\label{supp:fig:discrete_cr}
\end{figure}

The compression ratio of a single-axis compressed streak image can be calculated using Equation (\ref{supp:eq:compression_ratio_1D})\cite{Ma2021a}:
\begin{equation}
\label{supp:eq:compression_ratio_1D}
CR = \frac{N_{x}N_{y}N_{t}}{N_{x}(N_{y}+N_{t})}.
\end{equation}
The TACSI compression ratio, given by Equation (\ref{supp:eq:compression_ratio}), can be directly compared to Equation (\ref{supp:eq:compression_ratio_1D}) by discretizing the variables and adopting a rectangular geometry. Starting from the general form of the TACSI streak surface area in Equation (\ref{supp:eq:streak_surface}), it follows that: 
\begin{align}
\begin{split}
\label{supp:eq:discrete_streak_surface}
N_{s'} &= \frac{w_{\perp}t\sqrt{v_{o}^{2}+v_{s}^{2}}+s_{o}}{p^{2}} \\
       &= \frac{w_{\perp}}{p}t\sqrt{\biggl(\frac{v_{o}}{p}\biggl)^{2}+\biggl(\frac{v_{s}}{p}\biggl)^{2}}+\frac{s_{o}}{p^{2}} \\
       &= \frac{w_{\perp}}{p}\frac{v_{s}t}{p}\sqrt{r^{2}+1}+\frac{s_{o}}{p^{2}} \\
       &= \frac{w_{\perp}}{p}N_{t}\sqrt{r^{2}+1}+\frac{s_{o}}{p^{2}}.
\end{split}
\end{align}
The perpendicular width of the streak for a scene with a rectangular geometry is:
\begin{align}
\begin{split}
\label{supp:eq:discrete_perp_width}
\frac{w_{\perp}}{p} &= N_{x}sin\theta + N_{y}cos\theta \\
                    &= N_{x}sin\biggl(tan^{-1}\biggl(\frac{1}{r}\biggl)\biggl) + N_{y}cos\biggl(tan^{-1}\biggl(\frac{1}{r}\biggl)\biggl).
\end{split}
\end{align}
For convenience, $r$ is defined as the ratio between the object and streak speed: 
\begin{equation}
\label{supp:eq:streak_ratio}
r = \frac{v_{o}}{v_{s}},
\end{equation}
and the surface area of the rectangular scene is:
\begin{equation}
\label{supp:eq:discrete_surface_area}
\frac{s_{o}}{p^{2}} = N_{x}N_{y}.
\end{equation}

The discrete form of the TACSI compression ratio can now be written as:
\begin{equation}
\label{supp:eq:discrete_compression_ratio_2D}
CR = \frac{N_{x}N_{y}N_{t}}{N_{t}\biggl[N_{x}sin\biggl(tan^{-1}\biggl(\frac{1}{r}\biggl)\biggl) + N_{y}cos\biggl(tan^{-1}\biggl(\frac{1}{r}\biggl)\biggl)\biggl]\sqrt{r^{2}+1}+N_{x}N_{y}}.
\end{equation}

While the number of recovered image frames in a single-axis streak image will be limited by the height of the scene and the number of row pixels:
\begin{equation}
\label{supp:eq:single-axis_frame_limit}
N_{t} \leq P_{y}-N_{y},
\end{equation}
TACSI requires considering constraints along the horizontal and vertical axis of the imaging sensor:
\begin{equation}
\label{supp:eq:tacsi_frame_limit}
N_{t} \leq min \biggl\{ P_{y}-N_{y},\frac{P_{x}-N_{x}}{r} \biggl\}.
\end{equation}

\pagebreak

\section{Alternative Streak Methods}
\label{supp:sec:alternative_methods}

\begin{figure}[ht!]
\centering\includegraphics[width=13cm]{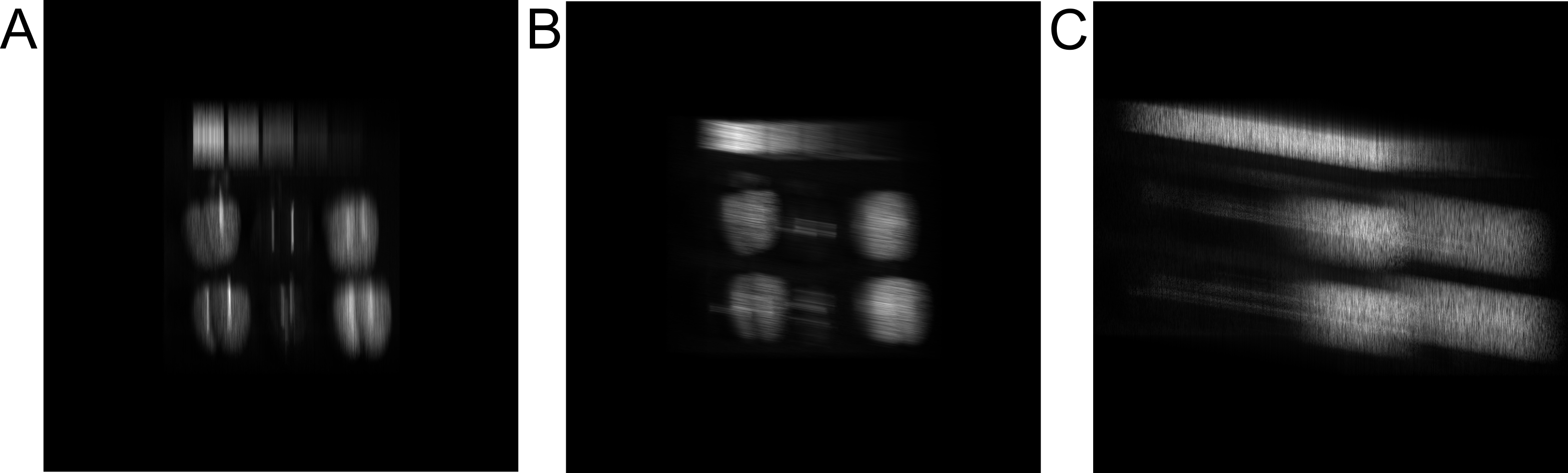}
\caption{Comparison of simulated compressed streak images generated with (A) conventional single-axis, (B) a two-axis galvo-scanner positioned between the CA and the camera, and (C) TACSI. The streak speed was held constant for all methods.}
\label{supp:fig:streak_method_comparison}
\end{figure}

Figure \ref{supp:fig:streak_method_comparison} shows simulated compressed streak images acquired using a hypothetical method in which a two-axis galvo-scanner is positioned between the coded aperture and the camera, conventional single-axis streak, and TACSI. The streak speed was held constant for all methods. The relationship between frame rate and streak speed is described by SI Appendix, Equation (\ref{supp:eq:streak_fps}). Both single-axis and TACSI ensure that temporal encoding is along the y-axis in Figure \ref{supp:fig:streak_method_comparison}A and \ref{supp:fig:streak_method_comparison}C. The two-axis galvo-scanner results in temporal encoding along the diagonal as seen in Figure \ref{supp:fig:streak_method_comparison}B.
Since a two-axis galvo-scanner would shear the mask along a diagonal, the concept of pixel pitch would need to be redefined. Determining this definition is outside of the scope of this manuscript. Furthermore, single-axis streak and the compressed streak image obtained with the two-axis galvo have identical motion blur because the object is not moving. The inability to decrease motion blur and compression ratio can be observed in Figure \ref{supp:fig:streak_method_comparison}B, where the coded aperture elements are not distinguishable.
For TACSI, because the streak speed does not depend on the speed of the object, it is possible to adjust the object speed while maintaining the frame rate. This allows the compression ratio and motion blur to be reduced simultaneously. For the compressed streak image in Figure \ref{supp:fig:streak_method_comparison}C, the object speed was 6-fold faster than the streak speed. It is important to note that the coded aperture elements in Figure \ref{supp:fig:streak_method_comparison}C can be distinguished. Similarly, no advantage can be gained by simply rotating the camera or galvo-scanner, and rotating the camera with respect to the CA would cause a mismatch between the CA elements and the camera pixels.

\end{appendices}


\bibliography{bibliography}

\end{document}